\documentclass[useAMS,usenatbib]{mn2e}
\usepackage[dvipdfmx]{graphicx,color}
\usepackage{amsmath}
\usepackage{textgreek}

\title[The contribution of bulk Comptonization to the soft X-ray excess in AGN]{The contribution of bulk Comptonization to the soft X-ray excess in AGN}
\author[J. Kaufman, O. M. Blaes, and S. Hirose]{J. Kaufman\thanks{E-mail:
jason.kaufman09@gmail.com (JK); blaes@physics.ucsb.edu (OMB)}, O. M. Blaes, and S. Hirose\\
Department of Physics, University of California, Santa Barbara, CA 93106, USA}
\begin{document}

\date{Accepted ---. Received ---; in original form ---}

\pagerange{\pageref{firstpage}--\pageref{lastpage}} \pubyear{2015}

\maketitle

\label{firstpage}

\begin{abstract}
Bulk velocities exceed thermal velocities for sufficiently radiation pressure dominated accretion flows. We model the contribution of bulk Comptonization to the soft X-ray excess in AGN. Bulk Comptonization is due to both turbulence and the background shear. We calculate spectra both taking into account and not taking into account bulk velocities using scaled data from radiation magnetohydrodynamic (MHD) shearing box simulations. We characterize our results with temperatures and optical depths to make contact with other warm Comptonization models of the soft excess. We chose our fiducial mass, $M = 2 \times 10^6 M_{\odot}$, and accretion rate, $L/L_{\rm Edd} = 2.5$, to correspond to those fit to the super-Eddington narrow line Seyfert 1 (NLS1) RE1034+396. The temperatures, optical depths, and Compton $y$ parameters we find broadly agree with those fit to RE1034+396. The effect of bulk Comptonization is to shift the Wien tail to higher energy and lower the gas temperature, broadening the spectrum. Observations of the soft excess in NLS1s can constrain the properties of disc turbulence if the bulk Comptonization contribution can be separated out from contributions from other physical effects, such as reflection and absorption.
\end{abstract}


\section{Introduction}
\label{sec_intro}
The soft X-ray excess in AGN spectra is the component below $1$keV that lies on top of the extrapolation of the best fitting 2-10keV power law \citep{sin85,arn85,vas14}. The dependence of effective temperature on mass and accretion rate in optically thick accretion disc models (\citealt{sha73}, hereafter SS73) is $T_{\rm eff} \sim (\dot{m}/M)^{1/4}$, where $\dot{m} = \dot{M}/\dot{M}_{\rm Edd}$. We therefore expect intrinsic disc emission to contribute to the soft excess most in narrow line Seyfert Is (NLS1), which are comparatively low mass ($\sim 10^6M_{\odot}$), near-Eddington sources. In the most luminous regions of NLS1 discs the temperature is greater than the hydrogen ionization energy, so electron scattering is the dominant opacity. The color temperature is therefore greater than the effective temperature, which augments the expected contribution to the soft excess in these sources. While the soft excess is particularly prominent in NLS1s, the expected disc contribution is insufficient to account for it (\citealt{don12}, hereafter D12). In broad line Seyferts, which are lower Eddington ratio sources, the intrinsic disc emission does not extend to high enough energies to contribute at all, and so in these sources the entire soft excess must originate elsewhere.

One class of models for the soft excess invokes warm Comptonization. In this picture, a warm ($kT_{\rm e} \sim 0.2$ keV) medium with moderate optical depth upscatters photons from a cool, optically thick disc. \cite{mag98}, for example, fit the soft excess of the broad line Seyfert 1 NGC 5548 with $kT_{\rm e} = 0.3$keV, $\tau = 30$. In this case, they pictured the medium as a transition region between the accretion disc and an inner hot geometrically thick flow. In other studies the medium is a warm layer above the inner regions of the disc. For example, \cite{jan01} fit the soft excess of the quasar PG 1211+143 with $kT_{\rm e} = 0.4$keV, $\tau = 10$.  \cite{dew07} fit two NLS1s, Ark 564 and Mrk 1044, with $kT_{\rm e} = 0.18$keV, $\tau = 45$, and $kT_{\rm e} = 0.14$keV, $\tau = 45$, respectively. \cite{jin09} fit the super-Eddington ($L/L_{\rm Edd} = 2.7$) NLS1 RXJ0136.9-3510 with $kT_{\rm e} = 0.28$keV, $\tau = 12$. \cite{meh11} fit the broad line Seyfert 1 Mrk 509 with $kT_{\rm e} = 0.2$keV, $\tau = 17$. More recently, D12 constructed the XSPEC model OPTXAGNF for the soft excess, which uses the disc spectrum at the outer coronal radius as the seed photon source and, for the purpose of energy conservation, models the warm medium as part of the disc atmosphere. D12 fit the super-Eddington ($L/L_{\rm Edd} = 2.4$) NLS1 RE 1034+396 with $kT_{\rm e} = 0.23$keV, $\tau = 11$. Since then, this model has been applied to several sources, such as the NLS1 II Zw 177 \citep{pal16}, for which they found $kT_{\rm e} \sim 0.2$keV, $\tau \sim 20$.

Warm comptonization models fit the spectra well, but the minimal variation of the fitted electron temperature with black hole mass and accretion rate (e.g. \citealt{gie04}) motivated alternative models based on discrete atomic features. In reflection models, photons from the hot ($\sim 100$ keV) corona are reflected and relativistically blurred by the inner regions of the accretion disc (e.g. \citealt{cru06}; \citealt{ros05}). In ionized absorption models, high velocity winds originating from the accretion disc absorb and reemit photons from the hot corona \citep{gie04}. While these models naturally predict the minimal variation in the soft excess temperature, they typically require extreme parameters to sufficiently smear the discrete atomic features on which they are based. Reflection models, for example, require near maximal spin black holes (e.g. \citealt{cru06}), and the original absorption models require unrealistically large wind velocities \citep{sch07}. More complex absorption models circumvent this difficulty, but they lack predictive power (e.g. \citealt{mid09}). Other proposed explanations for the soft excess include magnetic reconnection \citep{zho13} and Comptonization by shock heated electrons \citep{fuk16}. Because warm Comptonization, reflection, and absorption all fit the spectra adequately (e.g. \citealt{mid09}), solving this problem requires variability and multiwavelength studies (e.g. \citealt{meh11}; \citealt{vas14}). 

Because optically thick disc models predict that disc emission associated with NLS1s already extends into the soft X-rays, in these sources warm Comptonization could be due to modifications to the vertical structure that occur in this regime. For example, warm Comptonization may be due to turbulence in the disc (\citealt{soc04}; \citealt{kau16}, hereafter KB16), if bulk electron velocities exceed thermal electron velocities. For the alpha disk model\footnote{The $m_{\rm p}/m_{\rm e}$ factor in KB16 Eq. (1) should be flipped.} (SS73),
\begin{align}
\label{eq_ratio_1}
\frac{\left\langle v_{\rm turb}^2\right\rangle}{\left\langle v_{\rm th}^2 \right\rangle}
\sim \alpha \left(m_e\over m_{\rm p}\right) \left({P_{\rm rad}\over P_{\rm gas}}\right),
\end{align}
so we expect turbulent Comptonization to be important in the extreme radiation pressure dominated regime. Since the ratio of radiation to gas pressure increases with mass and accretion rate, turbulent Comptonization should be most relevant for supermassive black holes accreting near-Eddington, such as NLS1s.  In this regime, therefore, turbulent Comptonization could provide a physical basis for the construction of warm Comptonization models. By connecting the observed temperature and optical depth to the disc vertical structure, this could help solve the problem of the soft excess and also shed light on the properties of MHD turbulence. In broad line Seyferts, which have lower Eddington ratios, the ratio of radiation to gas pressure is too small for turbulent Comptonization to be significant, so if warm Comptonization is present it must originate elsewhere. In these sources, it is unlikely that warm Comptonization could be due to modifications to the intrinsic disc atmosphere physics, because the thermal spectrum falls off at energies significantly below the soft X-rays.

KB16 outlined the fundamental physical processes underlying bulk Comptonization by turbulence in accretion disc atmospheres. In this paper we model the effect of bulk Comptonization on disc spectra using data from radiation MHD simulations \citep{hir09}, including both turbulent Comptonization and Comptonization by the background shear. We parametrize this effect by temperature and optical depth in order to make contact with observations fit by other warm Comptonization models. In particular, we compare our results to the temperature and optical depth fit to RE 1034+396 (D12), a super-Eddington NLS1 with an unusually large soft excess. The structure of this paper is as follows. In section \ref{sec_modeling} we describe our model in detail. In section \ref{sec_results} we describe our results, and in section \ref{sec_discussion} we discuss them. Finally, we summarize our findings in section \ref{sec_Conclusion}.

\section{Modeling bulk Comptonization}
\label{sec_modeling}
\subsection{Overview}
\label{sec_overview}

In order to facilitate comparisons with warm thermal Comptonization models of the soft X-ray excess, we seek to characterize the contribution of bulk Comptonization with a temperature and an optical depth. To do this, we use data from radiation MHD shearing box simulations to compute spectra both including and excluding bulk velocities. Since our simulation data is limited, we use a scheme to scale data from a simulation run with a particular radius, mass, and accretion rate to different sets of these parameters. We describe this scheme in section \ref{sec_scalings}. In this work we use data from simulation 110304a, which is similar to simulations 1112a and 1226b \citep{hir09}, but has a lower surface density, $\Sigma = 2.5 \times 10^4$g cm$^{-2}$, which results in a higher radiation to gas pressure ratio. The parameters of interest for 110304a are given in Table \ref{table_sim_param}.

We calculate the spectrum at a given timestep using Monte Carlo post processing simulations. For this work, we chose the $140$ orbit timestep at random. The details of our Monte Carlo implementation of bulk Compton scattering are in Appendix \ref{sec_MCdetails}. To isolate the effect of the turbulence alone, we also calculate spectra without the background shear. To model an entire accretion disc we calculate spectra at multiple radii. We discuss our choice of radii in section \ref{sec_radii}. The flux obtained at a particular radius corresponds to an Eddington ratio. If our scaling scheme were perfect, the corresponding Eddington ratios at the other radii would be the same by construction. We correct for minor discrepancies by normalizing the other spectra so that their corresponding Eddington ratios are the same.

We transport the spectra computed with bulk velocities at multiple radii to infinity and superpose the results to obtain the final, observed spectrum. We choose a viewing angle of $60^{\circ}$. At this angle the gravitational redshift approximately cancels the Doppler blueshift (D12, \citealt{zha97}), which allows us to use a Newtonian transport code. We chose this method because  it is easy to include the propagation of error bars, but we verified that our results are unchanged when a fully relativistic Kerr spacetime transport code \citep{ago97} is used instead. 

The spectra computed without bulk velocities are used as seed photon sources for a warm Comptonizing medium characterized solely by a uniform temperature and optical depth. We implement this by solving the Kompaneets equation at each radius. We then transport the resultant spectra to infinity to obtain the observed spectrum. We fit the observed spectrum Comptonized by the warm medium to the observed spectrum computed with bulk velocities by adjusting the temperature and optical depth. We explore the effect of varying the outer radius, $r_{\rm cor}$, of the warm Comptonizing medium on the goodness of fit parameter, $\chi^2/\nu$, and select the radius for which this parameter is minimized.

To provide insight into the physics of bulk Comptonization, we also perform spectral calculations in which the simulation data are truncated at the effective photosphere and the emissivity is zero everywhere except in the cells at the base. Since we expect bulk Comptonization to be dominated by the contribution from photons emitted at the effective photosphere, we expect the resulting temperature and optical depth to be nearly unchanged. We discuss this point more in section \ref{sec_discussion_2}.

\subsection{Scalings for radiation MHD shearing box simulation data}
\label{sec_scalings}
\begin{figure}
\includegraphics[width = 84mm]{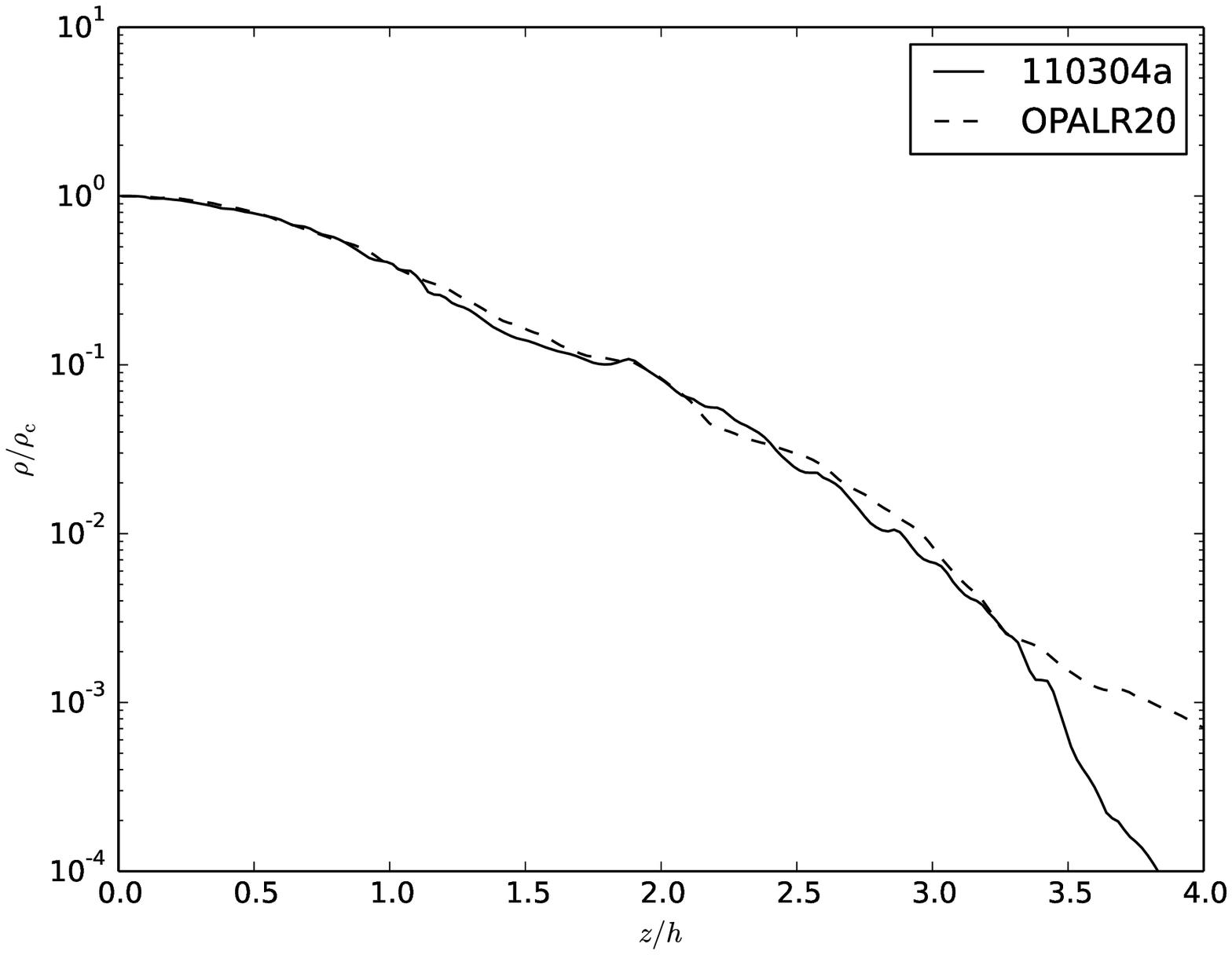}
\caption{Normalized shearing box density profiles at 140 orbits.}
\label{fig_density_140}
\end{figure}

\begin{figure}
\includegraphics[width = 84mm]{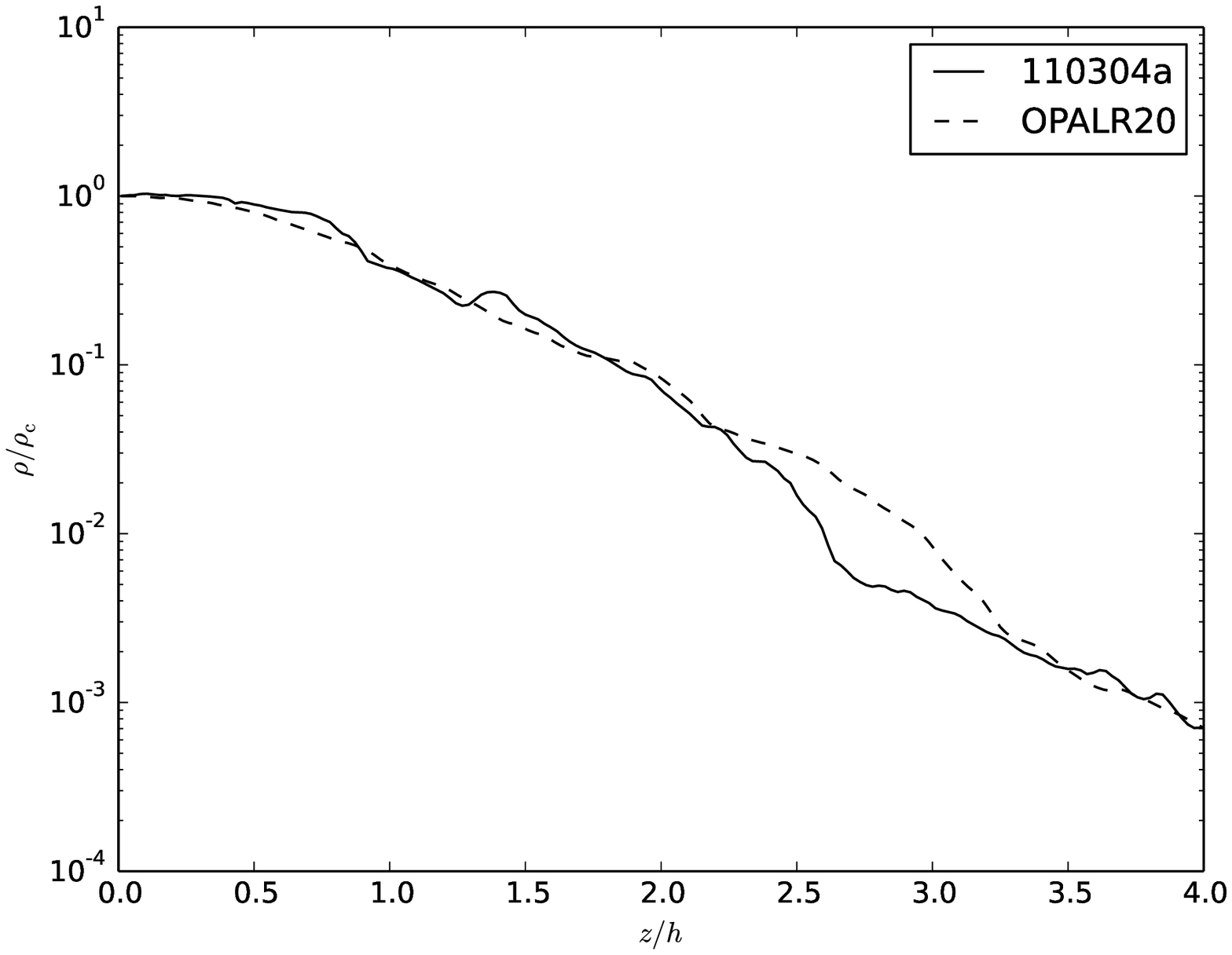}
\caption{Normalized shearing box density profiles at 180 orbits.}
\label{fig_density_180}
\end{figure}

\begin{figure}
\includegraphics[width = 84mm]{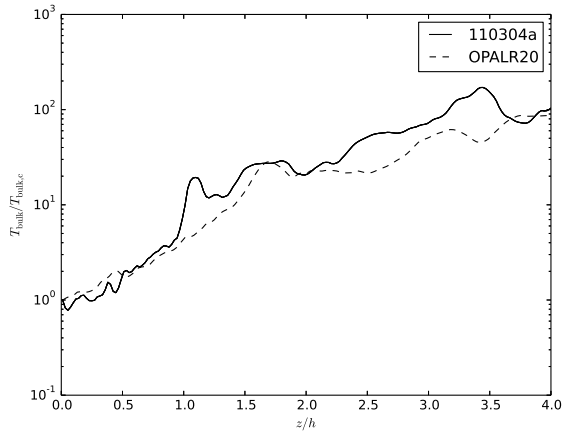}
\caption{Normalized shearing box bulk temperature profiles at 140 orbits.}
\label{fig_velocity_140}
\end{figure}

In this section we derive a scheme to scale data from a radiation MHD simulation run with a particular radius, mass, and accretion rate to a different set of these parameters. We first observe that the construction of an appropriate scheme is made possible by the fact that the density, temperature, and velocity profiles show considerable self-similarity across a wide range of simulation parameters. For example, in Figures  \ref{fig_density_140} and \ref{fig_velocity_140} we compare the density and bulk velocity profiles from the 140 orbit timestep of 110304a, which is the basis of this work, with those from a snapshot of OPALR20  \citep{jia16}, a simulation run in an entirely different regime (Table \ref{table_sim_param}). The bulk temperature is defined by $(3/2) k_{\rm B} T_{\rm bulk} = (1/2) m_e v^2$. Subscript ``${\rm c}$'' denotes midplane values. The variable $z$ is the distance from the midplane and the scale height $h$ is the value of $z$ for which $\rho / \rho_{\rm c} = 1/e$. The profiles nearly coincide, and even the discrepancy between the density profiles at large $z/h$ is likely just due to a temporary fluctuation at 140 orbits. At 180 orbits, for example, there is no discrepancy (Figure \ref{fig_density_180}). This self-similarity is perhaps an even more robust phenomenon than the difference in simulation parameters alone would indicate since the inclusion of the iron opacity bump in OPALR20 is a non-trivial effect. In particular, the thermal stability of OPALR20 depends on the inclusion of this effect \citep{jia16}, whereas it is now believed that the thermal stability in 110304a is a result of the narrow box size in the radial direction and is therefore artificial \citep{jia13}. Despite this caveat as well as the fact that the mass parameter for OPALR20 is closer to our regime of interest, we chose 110304a for this work because the photospheres are better resolved, a decisive advantage for the purpose of computing spectra.
\begin{table}
\caption{Shearing box simulation parameters}
\begin{tabular}{llll}
Simulation & $M/M_\odot$     & $L/L_{\rm Edd}$ & $r/r_{\rm g}$ \\
110304a    & 6.62            & 1.68            & 30 \\
OPALR20    & $5 \times 10^8$ & 0.03            & 40 \\ 
\end{tabular}
\label{table_sim_param}
\end{table}

Because of self-similarity, we primarily need to scale the midplane values for the profiles of interest and the scale height. Analogous to the derivation of the standard $\alpha$-disc scalings in the radiation pressure dominated regime (SS73), we derive scalings in terms of the shearing box surface density $\Sigma$, the vertical epicyclic frequency $\Omega_z$, and the shear $\partial_x v_y$. The integrated hydrostatic equilibrium equation for a density profile with scale height $h$ and midplane radiation pressure $P_{\rm c}$ is
\begin{align}
P_{\rm c} = \frac{1}{4}\Omega_z^2 \Sigma h.
\label{eq_hydro}
\end{align}
The thermal equilibrium equation, given the radiation flux $F$ and the midplane turbulent stress $\tau_{\rm c}$ is 
\begin{align}
F = (\partial_x v_y)\tau_{\rm c} h.
\label{eq_thermal}
\end{align}
The stress prescription is
\begin{align}
\bar{\tau} = \alpha \bar{P},
\end{align}
which for a profile that decays with scale height $h$ is equivalent to
\begin{align}
\tau_{\rm c} = \alpha P_{\rm c}.
\label{eq_stress}
\end{align}
The radiative diffusion equation with the opacity given by $\kappa$ is
\begin{align}
F = \frac{2cP_{\rm c}}{\kappa \Sigma}.
\label{eq_diff}
\end{align}
Eqs. (\ref{eq_thermal}), (\ref{eq_stress}), and (\ref{eq_diff}) give the scale height scaling:
\begin{align}
\left(\frac{h}{h_0}\right) = \left(\frac{\alpha}{\alpha_0}\right)^{-1} \left(\frac{\kappa}{\kappa_0}\right)^{-1} \left(\frac{\partial_x v_y}{\partial_x v_{y,0}}\right)^{-1} \left(\frac{\Sigma}{\Sigma_0}\right)^{-1}.
\label{eq_h}
\end{align}
Since we intend to scale to the lower mass ($\sim 10^6 M_\odot$), high Eddington ratio regime, the opacity remains dominated by electron scattering so we set $\kappa/\kappa_0 = 1$. Eqs. (\ref{eq_hydro}) and (\ref{eq_h}) give the midplane pressure scaling: 
\begin{align}
\left(\frac{P_{\rm c}}{P_{\rm c,0}}\right) = \left(\frac{\alpha}{\alpha_0}\right)^{-1}
\left(\frac{\kappa}{\kappa_0}\right)^{-1}
\left(\frac{\Omega_z}{\Omega_{z,0}}\right)^{2}
\left(\frac{\partial_x v_y}{\partial_x v_{y,0}}\right)^{-1}. 
\label{eq_pressure_c}
\end{align}
Below we will also need the flux scaling:
\begin{align}
\left(\frac{F}{F_0}\right) = \left(\frac{\alpha}{\alpha_0} \right)^{-1}
\left(\frac{\kappa}{\kappa_0}\right)^{-2} \left(\frac{\Omega_z}{\Omega_{z,0}} \right)^2 \left(\frac{\partial_x v_y}{\partial_x v_{y,0}}\right)^{-1}  \left(\frac{\Sigma}{\Sigma_0}\right)^{-1}.
\label{eq_flux_scale}
\end{align}
For the purpose of calculating spectra, the profiles of interest are the density, the gas temperature, the turbulent velocity, and the shear velocity. The midplane density is trivially given by
\begin{align}
\left(\frac{\rho_{\rm c}}{\rho_{\rm c,0}}\right) = \left(\frac{\Sigma}{\Sigma_0}\right) \left(\frac{h}{h_0}\right)^{-1}. 
\label{eq_density_c}
\end{align}
Since the gas temperature is coupled to the radiation temperature, the scaling for the midplane gas temperature follows directly from Eq. (\ref{eq_pressure_c}). To find the turbulent velocity scaling, we define $\beta$ as follows:
\begin{align}
\frac{1}{2} \left\langle \rho v^2 \right\rangle =  \beta \tau.
\end{align}
The midplane turbulent velocity scaling is then
\begin{align}
\frac{\left\langle v_{\rm c}^2 \right\rangle}{ \left\langle v_{\rm c,0}^2 \right\rangle} =& \left(\frac{\alpha}{\alpha_0}\right)^{-1} \left(\frac{\beta}{\beta_0}  \right)
\left(\frac{\kappa}{\kappa_0}\right)^{-2} \left(\frac{\Omega_z}{\Omega_{z,0}}\right)^2 \notag \\
&\left(\frac{\partial_x v_y}{\partial_x v_{y,0}} \right)^{-2}
\left(\frac{\Sigma}{\Sigma_0} \right)^{-2}.
\label{eq_v_c}
\end{align}
To test these scalings, we scale the midplane values and the scale height from 110304a to the simulation parameters of OPALR20 and then divide by the actual midplane values and the scale height in OPALR20 (Table \ref{table_midplane}). We assume $ \beta/\beta_0 = 1.$
Taking into account the empirical turbulent stress ratio $\alpha/\alpha_0 = 2.38$, we see that the resulting ratios are all near unity, and that our scalings therefore capture the essential physics in the shearing box. This is even more remarkable given that our scalings only take into account Thomson scattering and radiation diffusion, while the iron opacity bump and vertical advection are non-trivial effects in OPALR20. 

\begin{table}
\caption{Ratios of variables predicted using 110304a data to variables measured in OPALR20, taking into account $\alpha/\alpha_0 = 2.38$.}
\begin{tabular}{ll}
Variable & Ratio \\
$h_{\rm scaled}/h$ & 0.9 \\
$T_{\rm g,c,scaled}/T_{\rm g,c}$ & 1.0 \\
$T_{\rm bulk,c,scaled}/T_{\rm bulk,c}$ & 0.9 \\
\end{tabular}
\label{table_midplane}
\end{table}

The density and turbulent velocity profiles follow directly from Eqs. (\ref{eq_h}), (\ref{eq_density_c}), and (\ref{eq_v_c}), but the pressure profile, which determines the gas temperature profile, is non-trivial. The density profile is
\begin{align}
\rho\left(z\right) = \left(\frac{\Sigma}{\Sigma_0} \right) \left(\frac{h}{h_0} \right)^{-1} \rho_0\left(h_0 z/h\right).
\end{align}
The turbulent velocity profile is
\begin{align}
v(z) =& \left(\frac{\alpha}{\alpha_0}\right)^{-1/2} \left(\frac{\beta}{\beta_0}  \right)^{1/2}
\left(\frac{\kappa}{\kappa_0}\right)^{-1} \left(\frac{\Omega_z}{\Omega_{z,0}}\right) \notag \\
&\left(\frac{\partial_x v_y}{\partial_x v_{y,0}} \right)^{-1}
\left(\frac{\Sigma}{\Sigma_0} \right)^{-1}
v_0(h_0 z/h).
\end{align}
But scaling the radiation pressure profile by adjusting only the scale height and the overall normalization is too simplistic a scheme for the purpose of calculating spectra because near the photosphere the flux begins to free stream and is no longer carried by radiative diffusion. In such a scheme, therefore, the profile will be least accurate in the region that it is most important. This difficulty can be addressed by imposing a boundary condition at the photosphere. Inside the photosphere, 
\begin{align}
P_{\rm ph,in} \sim T_{\rm ph, in}^4 \sim \left(f_{\rm cor} T_{\rm ph,out}\right)^4 \sim f_{\rm cor}^4 F, 
\end{align}
where $f_{\rm cor}$ is determined by the physics at the photosphere. For example, if the opacity is dominated by coherent scattering and the boundary condition is imposed at the effective photosphere, then $f_{\rm cor} = f_{\rm col}$, the color correction. The scaling for $P_{\rm ph,in}$ is then
\begin{align}
 P_{\rm ph,in} = \left(\frac{f_{\rm cor}}{f_{\rm cor,0}} \right)^4 \left(\frac{F}{F_0} \right) P_{\rm ph,in,0}.
\label{eq_pressure_ph}
\end{align}
The simplest scheme that imposes this boundary condition is given by
\begin{align}
\label{eq_pressure_profile}
P(z) =& P_{\rm ph,in} + \left(\frac{P_{\rm c}}{P_{\rm c,0}}\right)\left(P_0\left(h_0z/h\right) - P_0\left(h_0 z_{\rm ph}/h\right)\right),
\end{align}
which we formally derive in Appendix \ref{sec_pressure_scale}. We recall that $P_{\rm c}/P_{\rm c,0}$ is given by Eq. (\ref{eq_pressure_c}). Since the pressure at the photosphere is always orders of magnitude smaller than the midplane pressure, we find that
\begin{align}
P(0) \approx \left(\frac{P_{\rm c}}{P_{\rm c,0}}\right) P_0\left(0\right),
\end{align}
so that this scheme is self-consistent. Inside the photosphere the gas temperature is coupled to the radiation temperature, so in this region the gas temperature profile is then given by
\begin{align}
T_{\rm g,in}^4(z) =& T_{\rm g,ph}^4 + \left(\frac{P_{\rm c}}{P_{\rm c,0}}\right)\left(T_{\rm g,0}^4 \left(h_0z/h\right) - T_{\rm g,0}^4 \left(h_0 z_{\rm ph}/h\right)\right),
 \label{eq_Tgas_profile_1}
\end{align}
where
\begin{align}
T_{\rm g,ph}^4 = \left(\frac{P_{\rm ph,in}}{P_{\rm ph,in,0}}\right)T_{\rm g,ph,0}^4.
\label{eq_Tgas_ph}
\end{align}
In order that the gas temperature profile be continuous, the scaling outside the photosphere is given by
\begin{align}
T_{\text{g,out}}^4\left(z\right) =& \left(\frac{P_{\rm ph,in}}{P_{\rm ph,in,0}}\right) T_{\text{g},0}^4\left(z_{\text{ph},0} + h_0 (z-z_{\text{ph}})/h\right).
\label{eq_Tgas_profile_2}
\end{align}
Finally, we also need the scaling for the shear velocity profile, which is trivially given by
\begin{align}
v_{\rm s}\left(x\right) = \left(\frac{\partial_x v_y}{\partial_x v_{y,0}}\right)\left(\frac{h}{h_0}\right) v_{\rm s,0}\left(h_0 x/h\right).
\label{eq_vshear_profile}
\end{align}
We define $z_{\rm ph}$ to be where the scattering optical depth $\tau_{\rm s} = 1$ (where subscript ``s" denotes scattering) and set $f_{\rm cor}/f_{\rm cor,0} = 1$. Near the photosphere magnetic pressure begins to play a major role in hydrostatic equilibrium (e.g. \citealt{bla07}), and near the effective photosphere the gas temperature begins to diverge from the radiation temperature, so we acknowledge that the assumptions underlying our scheme do not reflect the detailed physics in this region. But since our goal is only to calculate spectra, for optical depths $\tau_{\rm s} \ll 1$ the accuracy of this scheme is not important. We can assess the validity of this scheme in the region $\tau_{\rm s} \approx 1$ by comparing the flux from spectral calculations with the intended flux given by Eq. (\ref{eq_flux_scale}), or, equivalently, by comparing the corresponding Eddington ratios. In section \ref{sec_full}, we make this comparison for each set of scaling parameters we use and find that they generally agree to within 10$\%$. More importantly, we find that normalizing the spectra at different radii so that their corresponding Eddington ratios match has a neglible impact on the observed spectrum when contrasted with the discrepances between spectral calculations with and without bulk velocities. In other words, because the potential error is significantly less than the effect we are measuring, our scaling scheme is adequate.

These are the appropriate equations for scaling data to a different set of fundamental shearing box simulation parameters, in particular $\Omega_z$, $\partial_x v_y$, and $\Sigma$. If we substitute in Eq. (\ref{eq_flux_scale}) for $\Sigma$, we can alternatively regard $F$ as a fundamental parameter instead of $\Sigma$. Shearing box scalings in terms of $F$ are given in Appendix \ref{sec_shearing_scale}. This substitution is useful in order to scale to a different set of fundamental accretion disc parameters, since it is straightforward to express $F$ in terms of accretion disc radius, mass, and accretion rate. The scalings for $\Omega_z$, $\partial_x v_y$, and $F$ for both Newtonian and Kerr discs, allowing for a non-zero stress inner boundary condition, are given in Appendix \ref{sec_parameter_scale}. The final scalings for $\rho$, $T_{\rm g}$, $v$, and $v_{\rm s}$ in terms of fundamental accretion disc parameters are given in Appendix \ref{sec_final_scale}. We only use Kerr scalings for our spectral calculations, but the Newtonian scalings are potentially useful for the purpose of comparing with other works in which Newtonian parameters are used and also for developing physical intuition.

\subsection{Dependence of turbulent Comptonization on radius}
\label{sec_radii}
\begin{figure}
\includegraphics[width = 84mm]{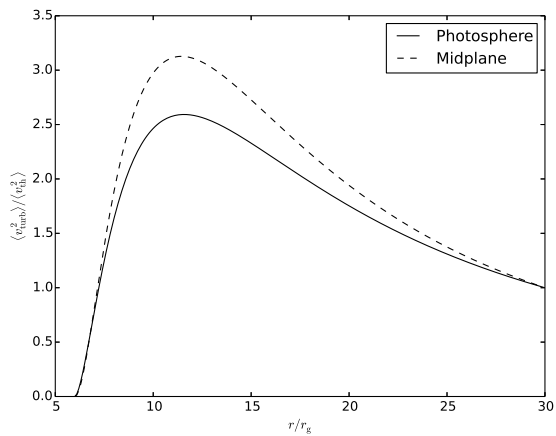}
\caption{Scaling for the relative magnitude of the turbulent velocity for $r_{\rm in} = 6 r_{\rm g}$, normalized to $r = 30 r_{\rm g}$.}
\label{fig_turb_comp_ratio}
\end{figure}

To characterize the contribution of turbulent Comptonization, we must model spectra at multiple radii. Our choice of radii is guided by the scaling of the ratio of bulk to thermal electron energies. We estimate this effect for a disc with no spin and a stress-free inner boundary condition with the Newtonian scalings in Appendix \ref{sec_final_scale}. The bulk velocity scaling is
\begin{align}
\left\langle v_{\rm turb}^2 \right\rangle \sim r^{-3}\left(1-\sqrt{r_{\rm in}/r}\right)^2.
\end{align}
The photosphere thermal velocity scaling is
\begin{align}
\left\langle v_{\rm th,ph}^2 \right\rangle  \sim r^{-3/4} \left(1-\sqrt{r_{\rm in}/r}\right)^{1/4}.
\end{align}
The scaling for the ratio of bulk velocity to thermal velocity at the photosphere is
\begin{align}
\label{eq_turb_ratio_1}
\frac{\left\langle v_{\rm turb}^2 \right\rangle}{\left\langle v_{\rm th,ph}^2 \right\rangle}
\sim r^{-9/4}\left(1-\sqrt{r_{\rm in}/r}\right)^{7/4}.
\end{align}
We also calculate the scaling for the ratio of bulk to thermal velocity using the midplane thermal velocity scaling, which is
\begin{align}
v_{\rm th,c}^2 \sim r^{-3/8}.
\end{align}
The scaling for the ratio is
\begin{align}
\label{eq_turb_ratio_2}
\frac{\left\langle v_{\rm turb}^2 \right\rangle}{\left\langle v_{\rm th,c}^2 \right\rangle} \sim r^{-21/8} \left(1-\sqrt{r_{\rm in}/r}\right)^2.
\end{align}
We plot Eqs. (\ref{eq_turb_ratio_1}) and (\ref{eq_turb_ratio_2}) in Fig. \ref{fig_turb_comp_ratio}, normalized to 30 gravitational radii. We expect that turbulent Comptonization will be most significant between 8 and 20 gravitational radii. We verify this assumption in section \ref{sec_results}. For our model we choose to compute spectra at 30, 20, 14, 11, 10, 9.5, 9.0, 8.5, and 7.5 gravitational radii. We also run simulations for spin $a = 0.5$, for which $r_{\rm in} = 4.2$. For these we compute spectra at 30, 20, 15, 12, 10, 8, 7, 6, 5.5, and 5 gravitational radii.

\section{Results}
\label{sec_results}

\begin{table*}
\caption{Simulation set independent variables}
\begin{tabular}{llllllll}
Set & Type & $M/M_{\odot}$ & $L/L_{\rm Edd}$ (target) & a & $\alpha/\alpha_0$ & $v_{\rm turb}$ & $v_{\rm shear}$\\
a & Full & $2 \times 10^6$   & 2.5 & 0   & 1 & Y & Y \\
a2 & Truncated, emissivity at base & $2 \times 10^6$   & 2.5 & 0   & 1 & Y & Y \\
b & Full & $2 \times 10^6$   & 2.5 & 0   & 1 & Y & N \\
b2 & Truncated, emissivity at base & $2 \times 10^6$   & 2.5 & 0   & 1 & Y & N \\
c & Full & $2 \times 10^6$   & 2.5 & 0   & 2 & Y & Y \\
c2 & Truncated, emissivity at base & $2 \times 10^6$   & 2.5 & 0   & 2 & Y & Y \\
d & Full & $2 \times 10^6$   & 2.5 & 0.5 & 1 & Y & Y \\
d2 & Truncated, emissivity at base & $2 \times 10^6$   & 2.5 & 0.5 & 1 & Y & Y \\
e & Full & $2\times 10^7$   & 2.5 & 0   & 1 & Y & Y \\
e2 & Truncated, emissivity at base & $2\times 10^7$   & 2.5 & 0   & 1 & Y & Y \\
\end{tabular}
\label{table_results_0}
\end{table*}

\begin{table*}
\caption{Results for full atmosphere spectral calculations}
\begin{tabular}{lllllllllllll}
Set & $M/M_{\odot}$ & $L/L_{\rm Edd}$ (target) & a & $\alpha/\alpha_0$ & $v_{\rm turb}$ & $v_{\rm shear}$ & $L/L_{\rm Edd}$ (observed) & $kT_{\rm e}$ (keV) & $\tau$ & $r_{\rm cor} (r_{\rm g})$ & $y_p$ & $\chi^2/\nu $ \\
a & $2 \times 10^6$   & 2.5 & 0   & 1 & Y & Y & 2.5 & $ 0.14 \pm 0.0067$ & $15 \pm 1.4$ & $20$ & $0.26$ & $1$ \\
b & $2 \times 10^6$   & 2.5 & 0   & 1 & Y & N & 2.5 & $ 0.18 \pm 0.056$ & $11 \pm 4.2$ & $14$ & 0.14 & 1.7 \\
c & $2 \times 10^6$   & 2.5 & 0   & 2 & Y & Y & 2.3 & $ 0.17 \pm 0.012$ & $17 \pm 1.8$ & $ 20 $ & 0.38 & 2.3 \\
d & $2 \times 10^6$   & 2.5 & 0.5 & 1 & Y & Y & 2.3 & $ 0.21 \pm 0.011$ & $12 \pm 0.82$ & 20 & 0.22 & 1.9 \\
e & $2\times 10^7$   & 2.5 & 0   & 1 & Y & Y & 2.1 & $0.081 \pm 0.0075$ & $24 \pm 4.1$ & 20 & 0.37 & 0.87 \\

\end{tabular}
\label{table_results_1}
\end{table*}

\begin{table*}
\caption{Results for truncated atmosphere spectral calculations with emissivity only at the base.}
\begin{tabular}{llllllllllll}
Set & $M/M_{\odot}$ & $L/L_{\rm Edd}$ (target) & a & $\alpha/\alpha_0$ & $v_{\rm turb}$ & $v_{\rm shear}$ & $kT_{\rm e}$ (keV) & $\tau$ & $r_{\rm cor} (r_{\rm g})$ & $y_p$ & $\chi^2/\nu $ \\
a2 & $2 \times 10^6$   & 2.5 & 0   & 1 & Y & Y & $ 0.14 \pm 0.0065$ & $16 \pm 1.4$ & $30$  & $0.26$ & $0.67$ \\
b2 & $2 \times 10^6$   & 2.5 & 0   & 1 & Y & N & $ 0.13 \pm 0.013$  & $12 \pm 2.5$ & $20$  & $0.15$ & $1.3 $ \\
c2 & $2 \times 10^6$   & 2.5 & 0   & 2 & Y & Y & $ 0.18 \pm 0.015$ & $14 \pm 1.4$ & $ 30 $ & 0.28 & 0.93 \\
d2 & $2 \times 10^6$   & 2.5 & 0.5 & 1 & Y & Y & $ 0.18 \pm 0.011$ & $14 \pm 1.2$ & $ 20 $ & 0.28 & 0.93 \\
e2 & $2\times 10^7$   & 2.5 & 0   & 1 & Y & Y & $ 0.074 \pm 0.0040 $ & $32 \pm 4.5$ & $20$ & 0.57 & 0.52 \\
\end{tabular}
\label{table_results_2}
\end{table*}

\begin{table*}
\caption{Goodness of fit of parameters derived from truncated atmosphere spectral calculations to observed spectra calculated with the full atmosphere.}
\begin{tabular}{llllll}
Set & $kT_{\rm e}$ (keV) & $\tau$ & $r_{\rm cor} (r_{\rm g})$ & $y_p$ & $\chi^2/\nu $ \\
a & $ 0.14$ & $16$ & $30$  & $0.26$ & 1.6 \\
b & $ 0.13$ & $12$ & $20$  & $0.15$ & 2.0 \\
c & $ 0.18$ & $14$ & $ 30 $ & 0.28 & 2.6 \\
d & $ 0.18$ & $14$ & $ 20 $ & 0.28 & 1.9 \\
e & $ 0.074$ & $32$ & $20$ & 0.57 & 1.1 \\
\end{tabular}
\label{table_results_3}
\end{table*}

We compute the contribution of bulk Comptonization to the soft X-ray excess and characterize our results with a temperature and optical depth. Our fiducial mass, $M = 2 \times 10^6 M_{\odot}$, and Eddington ratio, $L/L_{\rm Edd} = 2.5$, were chosen to correspond to those of the NLS1 source RE 1034+396 in D12 (Table \ref{table_obs}). Table \ref{table_results_1} summarizes our main results. The original (unscaled) simulation parameters for 110304a are listed in Table \ref{table_sim_param}. Each system is modeled by calculating spectra with and without the bulk velocities at the set of radii discussed in section \ref{sec_radii}. The target $L/L_{\rm Edd}$ is the Eddington ratio that would correspond to the observed flux at 30 gravitational radii if the scaling scheme were exact. The turbulent stress scaling is given by $\alpha/\alpha_0$. In all cases, $\Delta \epsilon = 0$ (Appendix \ref{sec_parameter_scale}), which imposes the stress-free inner boundary condition. The choices of whether or not to include turbulent and shear velocities in the spectral calculations with bulk velocities are indicated by $v_{\rm turb}$ and $v_{\rm shear}$, respectively. The Compton $y$ parameter is calculated from the fitted temperature and optical depth. To calculate $\chi^2/\nu$, we first correct for uncertainty in the overall normalization of the data point errors by normalizing them to the standard deviation calculated from the fit for set (a) (shown in Fig. \ref{fig_fit_a}). In section \ref{sec_full}, we discuss the results of each set. To provide physical insight into the physics of bulk Comptonization, we also perform spectral calculations in which the simulation data was truncated at the effective photosphere and the emissivity was set to zero everywhere except in the cells at the base. Table \ref{table_results_2} summarizes these results, which we discuss in section \ref{sec_truncated}. For clarity, in Table \ref{table_results_0} we list the independent variables for all simulation sets.

\subsection{Full spectral calculations}
\label{sec_full}
\begin{figure}
\includegraphics[width = 84mm]{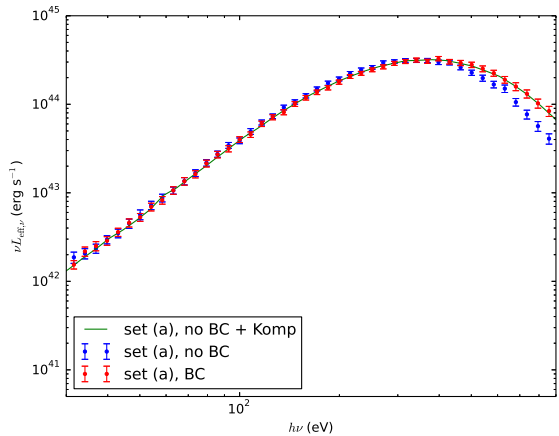}
\caption{Observed disc spectra computed for set (a). BC (bulk Comptonization) means bulk velocities were included. Komp means the zero bulk Comptonization spectrum from each radius for $r \leq r_{\rm cor}$ was passed through a warm Comptonizing medium with the parameters given in Table \ref{table_results_1}.}
\label{fig_fit_a}
\end{figure}

\begin{figure*}
\begin{tabular}{ll}
\includegraphics[width = 84mm]{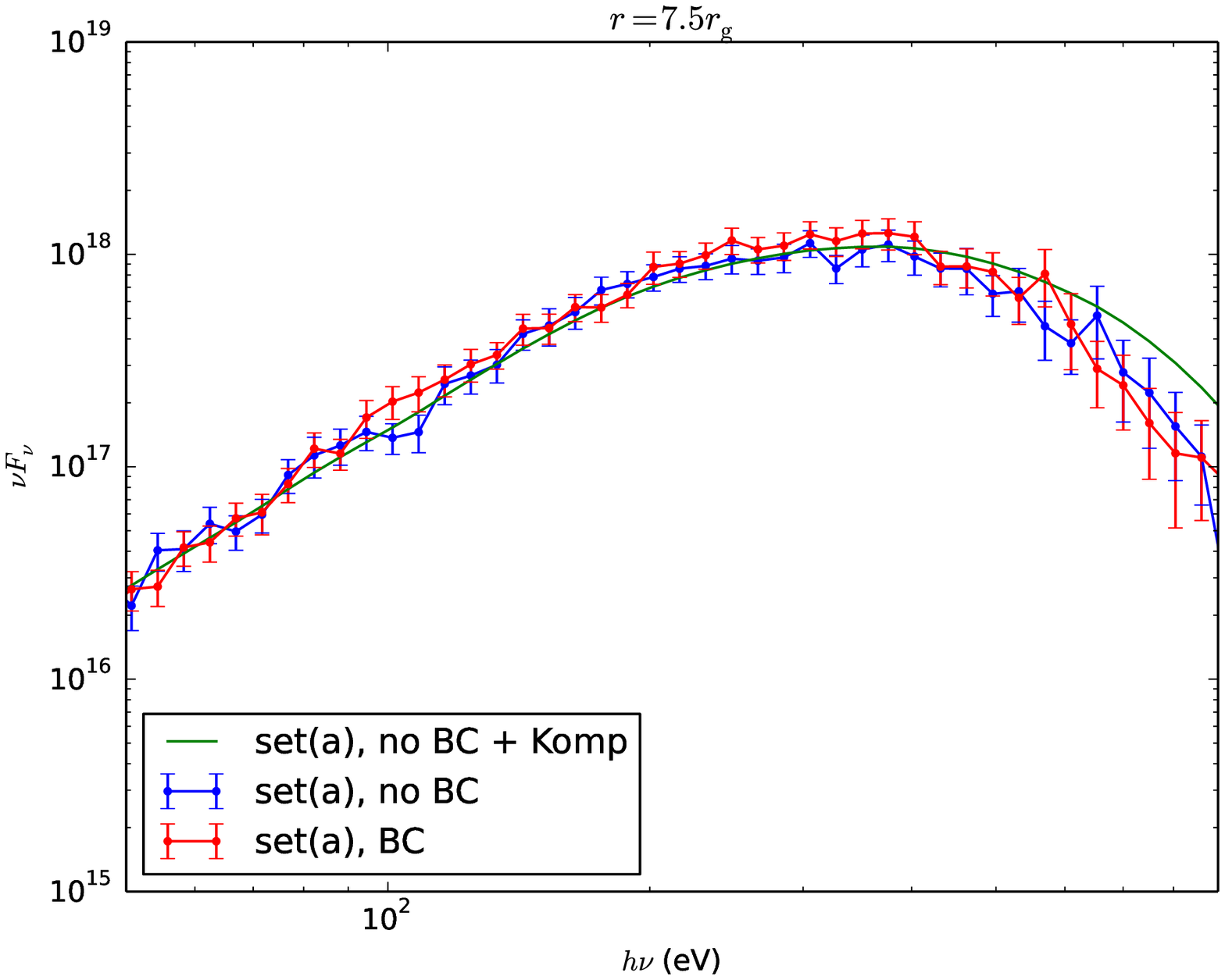} & \includegraphics[width = 84mm]{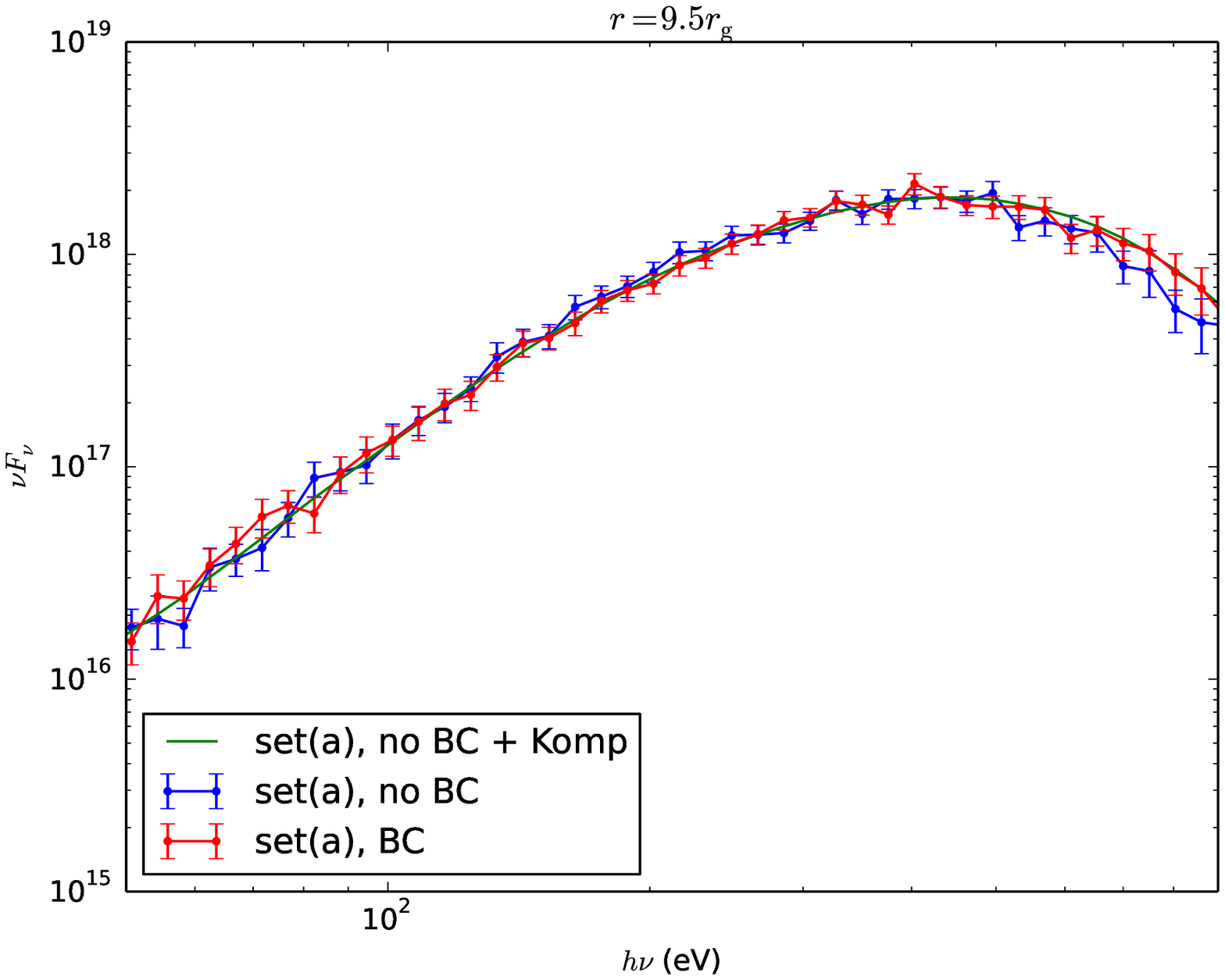} \\
\includegraphics[width = 84mm]{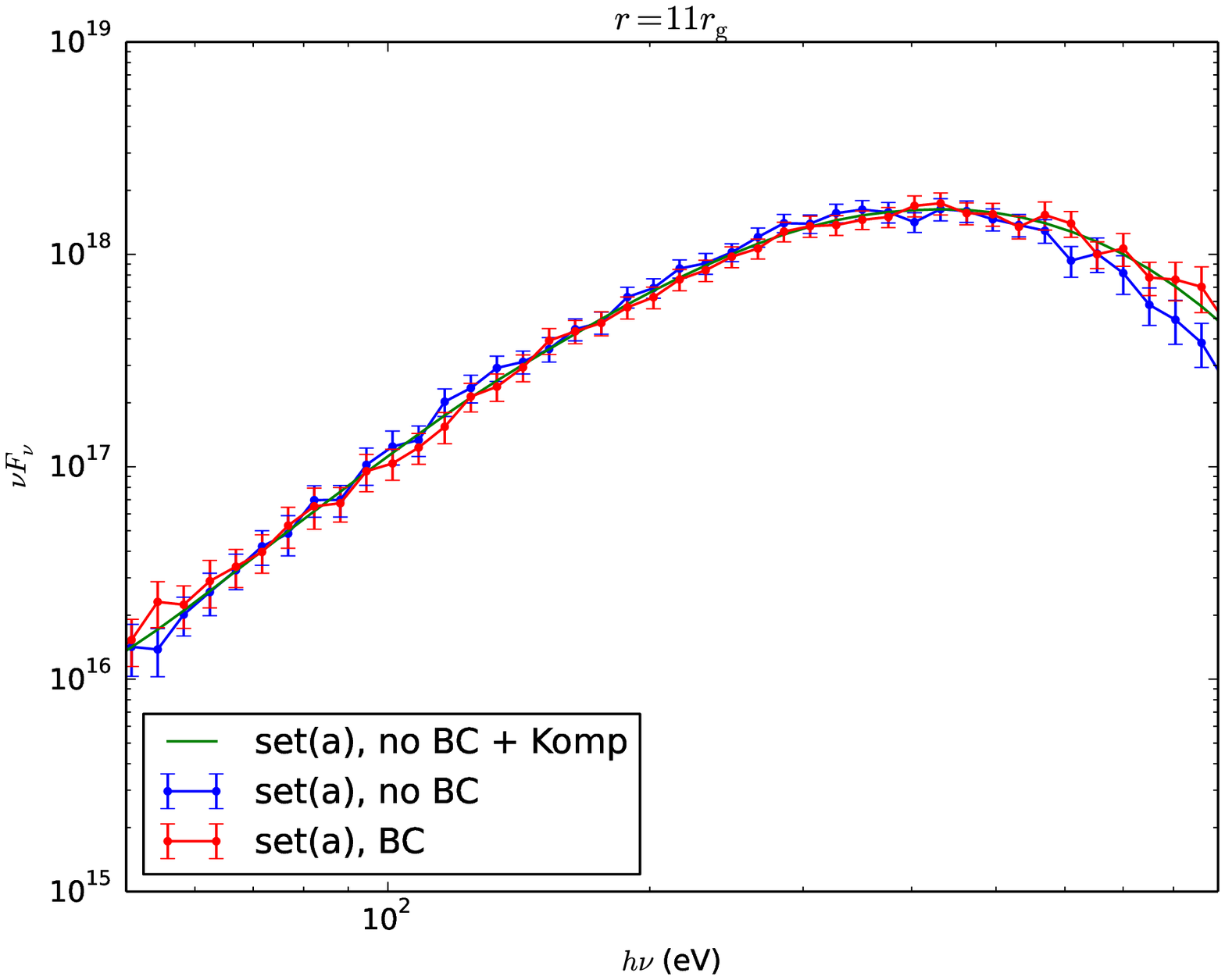} & \includegraphics[width = 84mm]{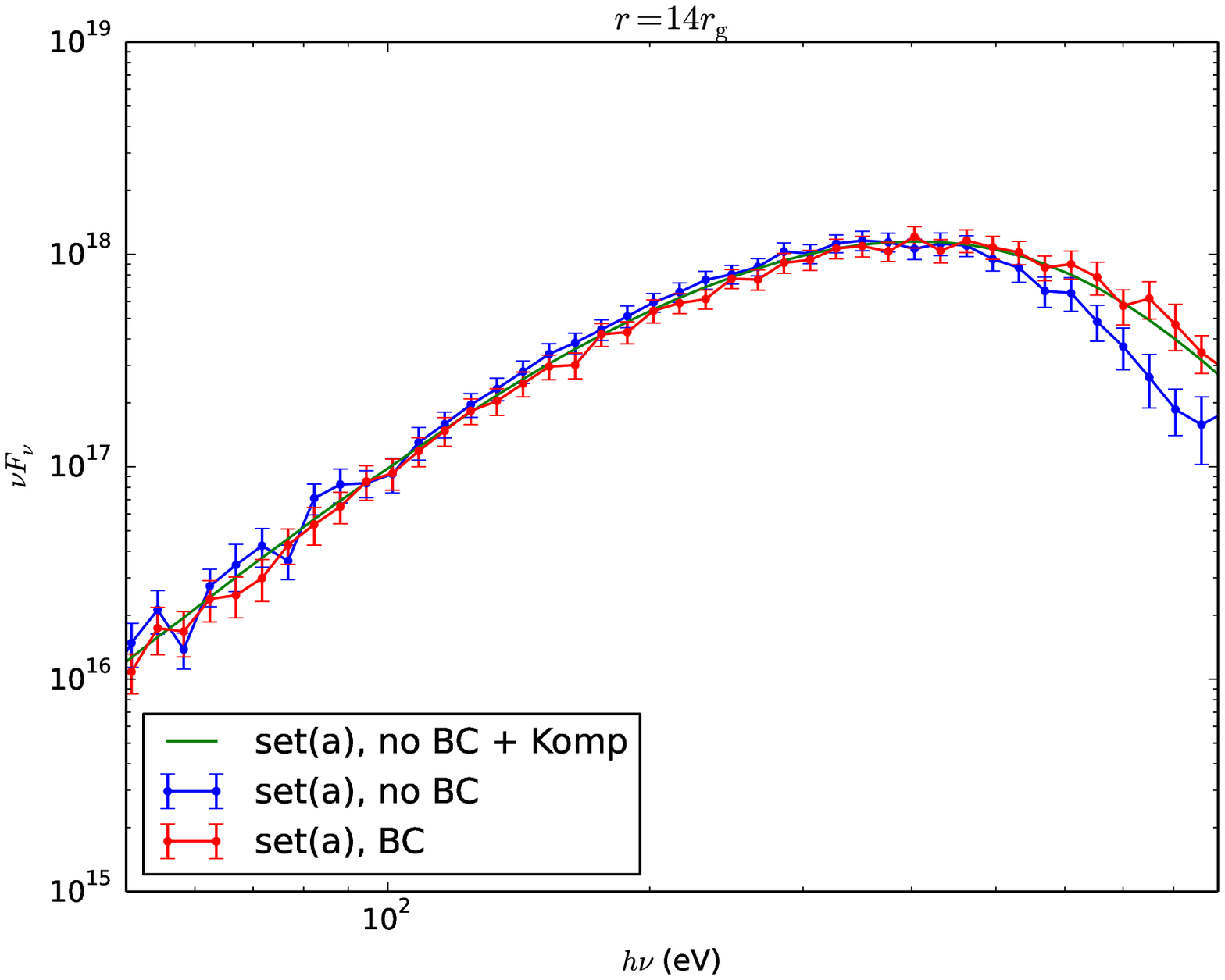} \\
\includegraphics[width = 84mm]{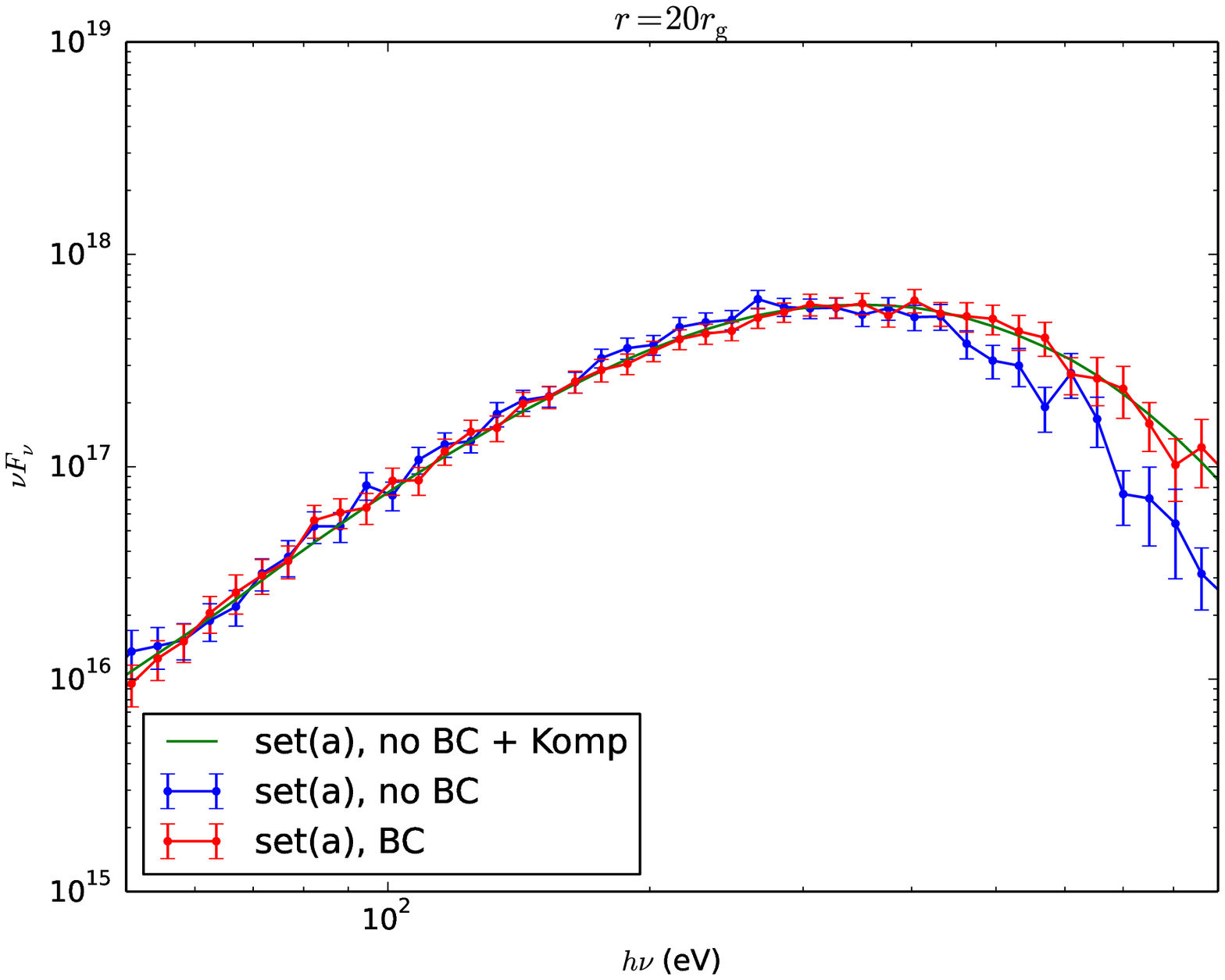} & \includegraphics[width = 84mm]{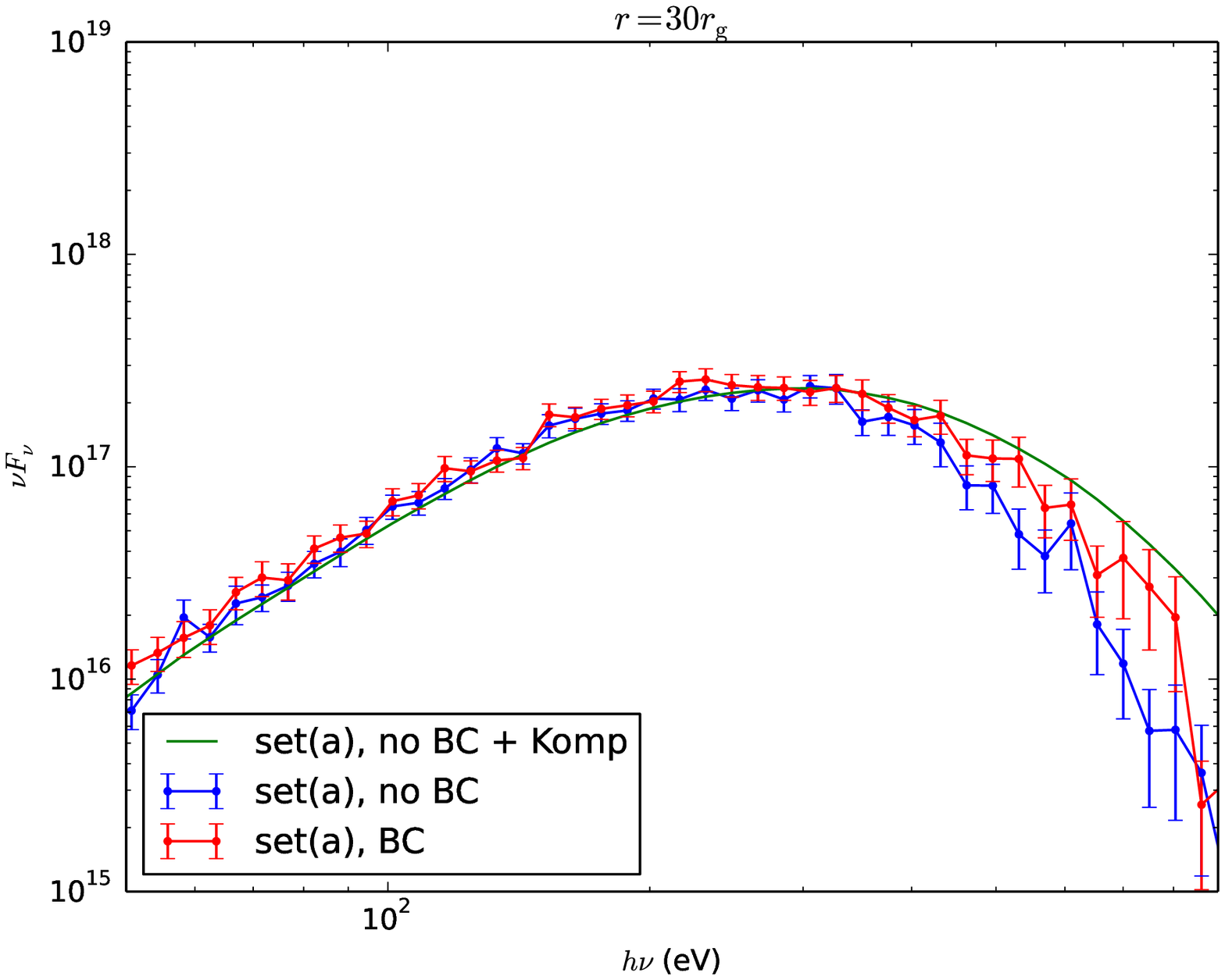} \\
\end{tabular}
\caption{Disc spectra at select radii, labeled at the top of each plot, computed for set (a). BC (bulk Comptonization) means bulk velocities were included. Komp means the zero bulk Comptonization spectrum was passed through a warm Comptonizing medium with the parameters given in Table \ref{table_results_1}.}
\label{fig_fit_a_r}
\end{figure*}

\begin{table}
\caption{Flux normalizations to the Eddington ratio at $r = 30 r_{\rm g}$ for set (a).}
\begin{tabular}{lll}
$r/r_{\rm g}$ & Flux norm (No BC) & Flux norm (BC) \\
30            & 1                 & 1\\
20            & 1.04              & 1.10\\
14            & 1.04              & 1.15\\
11            & 0.99              & 1.06\\
10            & 0.95              & 0.96\\
9.5           & 0.92              & 0.94\\
9.0           & 0.91              & 0.89\\
8.5           & 0.90              & 0.87\\
7.5           & 1.03              & 1.06\\
\end{tabular}
\label{table_flux_normalizations}
\end{table}

\begin{figure}
\includegraphics[width = 84mm]{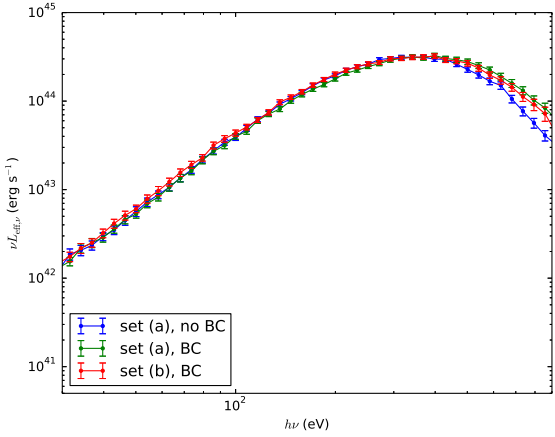}
\caption{Observed disc spectra computed for sets (a) and (b). BC (bulk Comptonization) means bulk velocities were included. Set (a) includes both turbulence and shear. Set (b) includes only turbulence.}
\label{fig_fit_b}
\end{figure}

\begin{figure}
\includegraphics[width = 84mm]{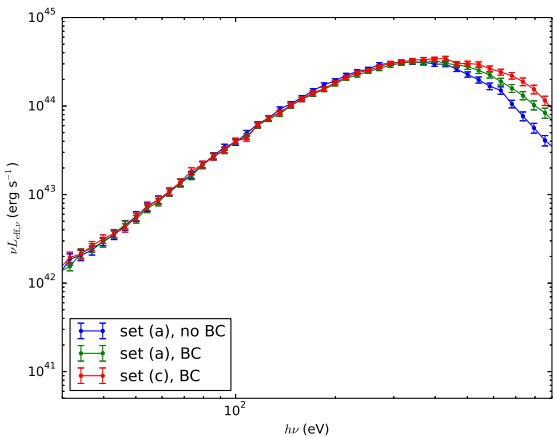}
\caption{Observed disc spectra computed for sets (a) and (c). BC (bulk Comptonization) means bulk velocities were included. For set (a) the turbulent stress scaling $\alpha/\alpha_0$ is 1. For set (c), $\alpha/\alpha_0 = 2$.}
\label{fig_fit_c}
\end{figure}

\begin{figure}
\includegraphics[width = 84mm]{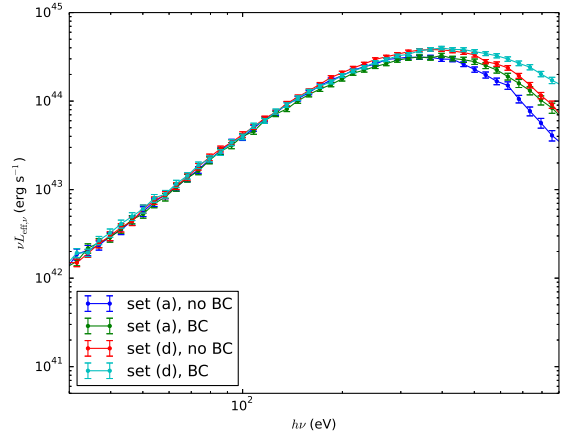}
\caption{Observed disc spectra computed for sets (a) and (d). BC (bulk Comptonization) means bulk velocities were included. For set (a), the spin parameter $a=0$. For set (d), $a = 0.5$.}
\label{fig_fit_d}
\end{figure}

The observed spectrum for set (a) computed with and without the bulk velocities along with the Kompaneets fit are shown in Figure \ref{fig_fit_a}. We see that the fit is excellent, which means that bulk Comptonization here is well modeled by thermal Comptonization with a fitted temperature and optical depth. We note that the observed $L/L_{\rm Edd}$ matches the target $L/L_{\rm Edd}$, which confirms that our scaling scheme is self-consistent. The required flux normalizations given the flux at 30 gravitational radii are given in Table \ref{table_flux_normalizations}. They hardly deviate from unity, which provides another check for the self-consistency of our scalings. In Fig. \ref{fig_fit_a_r} we show local spectra at multiple radii for set (a). We see that the spectra passed through the warm Comptonizing medium fit the spectra calculated with bulk velocities for $9.5 r_{\rm g} \leq r \leq 20 r_{\rm g}$, but overshoot them for $r = 7.5r_{\rm g}$ and $r = 30r_{\rm g}$. This confirms that bulk Comptonization is most significant in the region we expected it to be (section \ref{sec_radii}). Furthermore, this is consistent with the value we find for $r_{\rm cor}$, since we expect the best fit to be obtained when the Comptonizing medium is restricted to the region in which bulk Comptonization is most significant.

For set (b) we calculate spectra without the background shear to isolate the effect of turbulence. The resulting observed spectrum is plotted in Figure \ref{fig_fit_b}. We see that the spectrum computed without shear lies significantly closer to the spectrum computed with shear than to the spectrum computed without the bulk velocities. This indicates that bulk Comptonization is primarily due to turbulence, not shear.

For set (c) we test the robustness of our results by repeating spectral calculations with a different turbulent stress scaling ratio, $\alpha/\alpha_0 = 2$. For OPALR20 (section \ref{sec_scalings}), for example, $\alpha/\alpha_0 = 2.38$. The resulting observed spectrum is plotted in Figure \ref{fig_fit_c}. We see that although the observed spectrum computed with $\alpha/\alpha_0 = 2$ is Comptonized more than the spectrum computed with $\alpha/\alpha_0 = 1$, the effect is not huge. In particular, the fitted temperature and optical depth are only 21\% and 13\% higher, respectively. Since the turbulent velocity squared scales as $\alpha$ (Eq. \ref{eq_v_turb_AC}), one might expect that the fitted temperature would also scale as $\alpha$, but this neglects the contribution by shear as well as the fact that we are fitting the optical depth along with the temperature rather than holding the optical depth fixed. The magnitude of bulk Comptonization is better indicated by $y_{\rm p}$. From set (b) we see that for $\alpha/\alpha_0 = 1$, $y_{\rm p} = 0.14$ for turbulence alone. From sets (a) and (b) we infer that for $\alpha/\alpha_0 = 1$, $y_{\rm p} = 0.26 - 0.14 = 0.12$ for shear alone. We would expect, therefore, that for $\alpha/\alpha_0 = 2$, $y_{\rm p} = 2 \times 0.14 + 0.12 = 0.40$, which is very close to the fitted value $y_{\rm p} = 0.38$.

For set (d) we explore the effect of varying the spin parameter by setting $a = 0.5$. The resulting observed spectrum is plotted in Figure \ref{fig_fit_d}. As expected, the original spectra computed without bulk velocities are hotter and more luminous for the higher spin parameter since the accretion efficiency is higher. But the effect of bulk Comptonization is comparable. The fitted temperature is slightly higher, but the fitted optical depth is slightly lower, leading to an effect that is nearly the same.

Finally, for set (e) we use a higher mass, $M = 2\times 10^7 M_{\odot}$. The fitted temperature is lower, consistent with the dependence of overall accretion disc temperature on mass. But the larger value of $y_{\rm p}$ indicates that the effect of bulk Comptonization on the spectrum is greater. This is consistent with Eq. (\ref{eq_ratio_1}), since the ratio of radiation to gas pressure increases with mass (SS73). 

\subsection{Truncated atmosphere spectral calculations with emissivity only at the base}
\label{sec_truncated}

\begin{figure}
\includegraphics[width = 84mm]{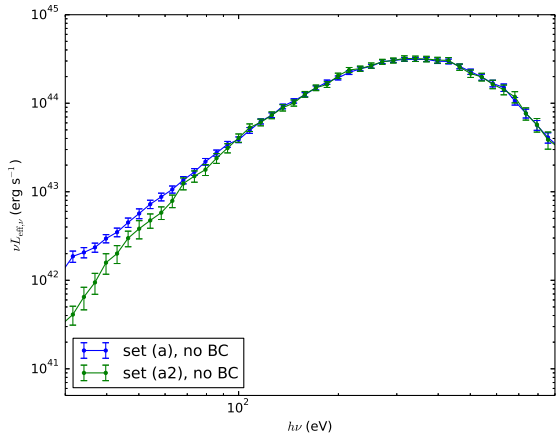}
\caption{Observed disc spectra computed for sets (a) and (a2). In set (a2), the atmosphere is truncated at the effective photosphere and the emissivity is zero everywhere except at in the cells at the base.}
\label{fig_fit_a2_1}
\end{figure}

\begin{figure}
\includegraphics[width = 84mm]{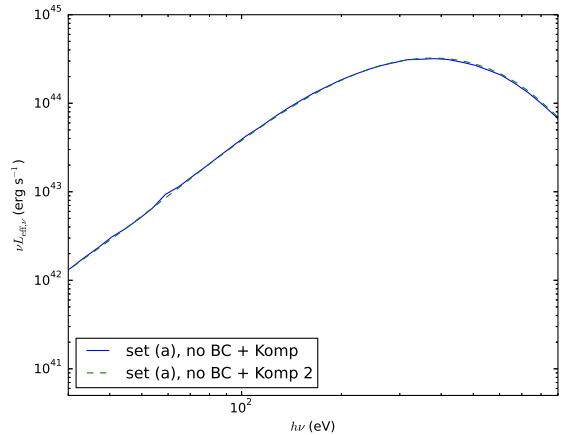}
\caption{Observed disc spectra computed for set (a). BC (bulk Comptonization) means bulk velocities were included. Komp means the zero bulk Comptonization spectrum from each radius for $r \leq r_{\rm cor}$ was passed through a warm Comptonizing medium with the parameters given in Table \ref{table_results_1}. For Komp 2 the parameters used are those fit to set (a2), given in Table \ref{table_results_2}.}
\label{fig_fit_a2_2}
\end{figure}

We expect that bulk Comptonization is predominantly explained by the Comptonization of photons emitted at the effective photosphere. We discuss this in detail in section \ref{sec_discussion_2}. To test this picture, we repeat spectral calculations with the parameters given in Table \ref{table_results_1} but truncate the atmosphere at the effective photosphere and set the emissivity to zero everywhere except in the cells at the base. Table \ref{table_results_2} summarizes these results.

For these calculations the observed spectra are different, but we expect the effect of bulk Comptonization on the observed spectra to be nearly unchanged. For example, the spectra computed without velocites for sets (a) and (a2), normalized to the total flux of (a), are plotted in Fig. \ref{fig_fit_a2_1}. The spectra coincide at high energies and diverge at low energies since photons emitted from lower temperature regions are omitted in (a2). But the fitted temperatures and optical depths for corresponding sets are very similar, which supports our picture of bulk Comptonization.

For sets (a) to (e), we also pass the spectra computed without the bulk velocities through a warm Comptonizing medium with the temperatures and optical depths fit to sets (a2) to (e2), respectively, and see whether the results fit the spectra computed with the bulk velocities. For each case we calculate $\chi^2/\nu$ to assess the goodness of fit and list the results in Table \ref{table_results_3}. In Fig. \ref{fig_fit_a2_2} for set (a) we plot the observed spectrum obtained by this procedure as well as the original fit. We see that the two curves nearly coincide and note that the corresponding values of $\chi^2/\nu$ differ by 0.6. For the other pairs of sets the corresponding values of $\chi^2/\nu$ differ by even less, which again confirms our expectation that bulk Comptonization is due to the Comptonization of photons emitted at the effective photosphere.

\section{Discussion}
\label{sec_discussion}

\begin{table*}
\caption{Fits to observed NLS1s}
\begin{tabular}{llllllll}
Source & Model & Reference & $M/M_{\odot}$ & $L/L_{\rm Edd}$ & $kT_{\rm e}$ (keV) & $\tau$ & $y_{\rm p}$\\ 
RE1034+396 & OPTXAGNF & D12 & $1.9 \times 10^6$ & 2.4      & $0.23 \pm 0.03$  & $11 \pm 1$ & 0.22\\ 
\end{tabular}
\label{table_obs}
\end{table*}

\subsection{Comparison with RE1034+396}
\label{sec_discussion_1}
In NLS1s the Wien tail of the intrinsic disc spectrum contributes to the soft excess (D12). Bulk Comptonization increases the contribution to the soft excess by shifting the Wien tail to higher energy. Since bulk Comptonization increases with accretion rate, we expect this contribution to be greatest in near and super-Eddington sources. In broad line Seyferts, the ratio of radiation to gas pressure is too low for bulk Comptonization to be significant. We compare our results to the analysis by D12 of RE1034+396, a super-Eddington NLS1 with an unusually large soft excess. This analysis is summarized in Table \ref{table_obs}. The comparison is appropriate because the mass and Eddington ratio we chose for our spectral calculations correspond to those fit to RE1034+396, though we do note that our model for Comptonization is more detailed than the one in D12.\footnote{In particular, in D12 the photon spectrum passed through the warm Comptonizing medium is given by the spectrum at $r_{\rm cor}$ and only the overall normalization varies with radius. This choice was made to minimize computation time.} We see that the Compton $y$ parameter, $y_{\rm p} = 0.22$, which characterizes the overall impact of Comptonization on the spectrum, is remarkably similar to the values we found. The fitted temperature and optical depth are also similar to our values. It may be, therefore, that the soft excess is unusually large in this system because of the contribution of bulk Comptonization.

A soft excess is also present in less luminous AGN for which bulk Comptonization is unlikely to be significant, and in general it seems that no single physical effect can fully explain the soft excess in all AGN. Until the contribution to the soft excess by other proposed mechanisms such as reflection and absorption are better understood, it will be difficult to tease out the contribution of bulk Comptonization. But our calculations show that if this can be done then observations of the soft excess can be used to constrain properties of the turbulence as well as other disc parameters.

\subsection{Physical interpretation of results}
\label{sec_discussion_2}
Comptonization of photons by bulk motions is due to effects both first and second order in velocity (KB16). The Kompaneets equation, which describes thermal Comptonization, cannot be used to describe first order effects, but KB16 showed that for incompressible motions in a periodic box with an escape probability, it does capture second order effects. The Kompaneets temperature for bulk velocities, which is a function of the photon mean free path, is given by
\begin{align}
k_{\rm B} T_{\rm Komp} = \frac{-2\lambda_{\rm p}m_{\rm e}c}{E} \left\langle P^{ij} \left(\partial_i v_j + \partial_j v_i \right) \right\rangle,
\end{align}
where $E$ is the radiation energy density and $P^{ij}$ is the radiation pressure tensor. Note that only the traceless part of the pressure tensor, which is the shear stress, contributes since this result assumes incompressible motions. We see that the Kompaneets temperature for bulk velocities is proportional to the stress multiplied by the strain rate, which is just the viscous dissipation of bulk motions by the photons. A stratified disc atmosphere is more complex than a periodic box, but the Kompaneets equation may still adequately describe second order effects. 

For our spectral calculations bulk Comptonization is well described by the Kompaneets equation, which suggests that second order effects, not first order effects, are dominant. This may be because MRI turbulence is incompressible and first order effects vanish for incompressible, but not compressible, turbulence (KB16). On the other hand, the photosphere regions are magnetically dominated and show considerable compressible motions because of the Parker instability \citep{bla07}, so it seems more likely that first order effects average out.

Assuming second order effects are dominant, we can gain physical insight into the fitted temperatures and optical depths by considering the dependence of the shear stress, $P^{ij}$, on the photon mean free path. The stress is largest when the photon mean free path is long relative to the maximum turbulence wavelength, and is proportional to the square of the photon mean free path when it is small (KB16). Therefore, Comptonization is only significant in the region near enough to the photosphere that the photon mean free path is comparable to the maximum turbulence wavelength. The resulting Comptonization temperature and optical depth should be the same for all photons emitted below this region. Inside this region, on the other hand, photons emitted nearer the photosphere should have comparatively larger Comptonization temperatures and smaller optical depths. For real disc atmospheres, which are stratified in (gas) temperature, photons contributing to the spectral peak are predominantly emitted at the effective photosphere, which for modest turbulence should be below the region where bulk Comptonization is significant. We therefore expect the resulting Comptonization temperature and optical depth to be unchanged when we truncate the atmosphere at the effective photosphere and set the emissivity equal to zero everywhere except at the base. Our findings confirm this. This is also useful because these spectral calculations run much faster which allows for a more efficient exploration of the disc parameter space.

\subsection{Self-consistency of results with shearing box simulations}
\label{sec_discussion_3}
We see that when bulk velocities are included in spectral calculations, the observed spectrum is shifted to higher energy. In particular, the Wien tail is shifted right. While this allows us to characterize bulk Comptonization with a temperature and optical depth as a function of accretion disc parameters, to determine the actual impact on disc spectra we must consider whether our spectral calculations are consistent with the underlying shearing box simulations on which they are based.

In section \ref{sec_discussion_2} we showed that bulk Comptonization here is predominantly an effect that is second order in velocity, but the underlying shearing box simulations \citep{hir09} do not include this effect because the flux limited diffusion approximation is used (KB16). Therefore, according to this picture we expect the spectral calculations without the bulk velocities to be consistent with the flux found in the underlying shearing box simulation. In order to determine the effect of including the bulk velocities on the resulting spectra we must take into account the back reaction on the vertical structure. Since adding in bulk Comptonization without modifying the vertical structure increases the flux and violates energy conservation, including this effect in the underlying shearing box simulation would lower the gas temperature until energy conservation is restored. Therefore, for significant Comptonization, while the Wien tail shifts to the right, the spectral peak shifts to the left. The overall effect, therefore, is to broaden the spectrum. In practice, if the Comptonization temperature is only slightly higher than the gas temperature then the spectrum will still be broadened but without an obvious leftward shift of the spectral peak. 

Because the decrease in gas temperature as well as other changes in the vertical structure may then affect bulk Comptonization, in theory the two should be calculated self-consistently. Another complicating factor is vertical advection of radiation, a velocity dependent effect that increases the number of photons emitted without affecting their energies, which also impacts energy conservation. But as long as bulk Comptonization is a perturbative effect, our fundamental results should hold: Bulk Comptonization broadens the spectrum by lowering the gas temperature and shifting the Wien tail to higher energy such that the total energy is conserved, and the characteristic temperatures and optical depths are given by Table \ref{table_results_1}. Furthermore, our method can be used to explore how bulk Comptonization scales with different parameters such as the mass, accretion rate, spin, turbulent stress scaling, and  boundary condition at the innermost stable circular orbit.

\subsection{Bulk Comptonization by the background shear}
\label{sec_discussion_4}
Our results suggest that Comptonization by bulk motions is predominantly due to turbulence, not shear. But since Comptonization by shear is not negligible, here we consider how it differs from Comptonization by turbulence, both in its potential effect on spectra and on the disc vertical structure.

From the perspective of total energy conservation, bulk Comptonization by the background shear at a given radius should have the same impact on the spectrum as turbulent Comptonization. It should shift the Wien tail to the right and decrease the gas temperature, broadening the spectrum. This is because the effective temperature for a steady state disc at a given radius is strictly fixed by the mass, mass accretion rate, and radius.

But Comptonization by the background shear plays a completely different role in the disc equations than Comptonization by turbulence. For the latter, the stress on the mean fluid flow is still entirely determined by MRI turbulence. For the $\alpha$ prescription, for example, the value of $\alpha$ is still set by the saturation level of the magnetic field and is therefore presumably unchanged. But Comptonization by shear is an additional stress on the mean fluid flow, and would therefore presumably increase $\alpha$. Since Comptonization by shear, at least in the regimes we have explored here, has only a perturbative effect on the spectrum, we expect any increase in $\alpha$ to be small. This is physically intuitive since dissipation by shear can be significant only near the photosphere where the mean free path is larger (see section \ref{sec_discussion_2}), whereas dissipation by MRI turbulence is significant throughout the body of the disc.

An interesting consequence of the difference between Comptonization by turbulence and background shear is that they have different effects on the total flux emitted from a shearing box. In a shearing box the density, not the radius, is fixed, and the flux depends on $\alpha$ (Eq. \ref{eq_flux_scale}). For Comptonization by turbulence, unless the MRI is affected, $\alpha$ is unchanged, and the gas temperature must decrease so that the flux is unchanged. But for Comptonization by the background shear, an additional source of stress on the mean flow is present, which modifies $\alpha$ and allows the flux to change. According to Eq. (\ref{eq_flux_scale}) we would ironically expect the flux to decrease rather than increase, but we should not take this prediction seriously. Comptonization by bulk motions is only significant near the photosphere where predominantly magnetic pressure, not radiation pressure, supports the atmosphere, so a small perturbation to $\alpha$ confined to this region cannot be treated self-consistently by the standard $\alpha$ disc equations.

Of course, in practice it is not well understood what determines $\alpha$; it is possible that even Comptonization by turbulence indirectly affects $\alpha$. It is also possible that Comptonization by the background shear indirectly decreases the saturation level of the magnetic field so that the net effect is to leave $\alpha$ unchanged. Our point is that Comptonization by turbulence and Comptonization by the background shear play different roles in the disc equations and therefore potentially have different effects on the vertical structure.

\section{Summary}
\label{sec_Conclusion}
We modeled the contribution of bulk Comptonization to the soft X-ray excess in AGN. To do this, we calculated disc spectra both taking into account and not taking into account bulk velocities with data from radiation MHD simulations. Because our simulation data was limited, we developed a scheme to scale the disc vertical structure to different values of radius, mass, and accretion rate. For each parameter set, we characterized our results by a temperature and optical depth in order to facilitate comparisons with other warm Comptonization models of the soft excess. We chose our fiducial mass, $M = 2 \times 10^6 M_\odot$, and accretion rate, $L/L_{\rm Edd} = 2.5$, to correspond to the values fit by D12 to the super-Eddington narrow-line Seyfert 1 RE1034+396, which has an unusually large soft excess. Our principle results are as follows.

For zero spin, when Comptonization by both turbulence and the background shear are included, the Compton $y$ parameter we find, $y_{\rm p} = 0.26$, is close to that found by D12 for RE1034+396, $y_{\rm p} = 0.22$. The temperature we find is a bit lower ($kT  = 0.14$keV vs. $kT = 0.23$keV), but the optical depth is higher ($\tau = 15$ vs. $\tau = 11$). For spin $a = 0.5$, the correspondence is remarkable; we find $y_{\rm p} = 0.22$, $kT = 0.21$keV, and $\tau = 12$. We find that bulk Comptonization is primarily due to turbulence, not the background shear (Fig. \ref{fig_fit_b}). Both the fitted temperature and optical depth increase moderately when we double the turbulent stress scaling to $\alpha/\alpha_0  = 2$. When we increase the mass, the fitted temperature decreases, but the $y$ parameter increases. This indicates that bulk Comptonization is more significant, which we expect since the ratio of electron thermal to bulk velocities depends on the ratio of radiation to gas pressure (Eq. \ref{eq_ratio_1}), which in turn scales with mass (SS73). Our results are given in Table \ref{table_results_1}.

To enforce energy conservation, the impact of bulk Comptonization on disc spectra is to shift the Wien tail to the right while simultaneously lowering the gas temperature, broadening the spectrum. Since we find that bulk Comptonization is well described by the Kompaneets equation, this suggests that it is predominantly an effect second order in velocity (KB16). Knowledge of this is important for self-consistently resolving bulk Comptonization in radiation MHD simulations, since common closure schemes such as flux limited diffusion do not include this effect (KB16).

The soft excess in general is unlikely due to a single physical mechanism. 
Other contributing effects, such as reflection and absorption, must be better understood to make precise comparisons of predictions by models of bulk Comptonization with observations. But the fact that our results, based simply on the most naive scalings, are in agreement with observations suggests that at least in the super-Eddington NLS1 regime bulk Comptonization may play a significant role in producing the soft X-ray excess. If so, observations of the soft excess can be directly tied to the properties of MHD turbulence as well as fundamental disc parameters.

\section*{Acknowledgements}
We thank Yan-Fei Jiang for providing us with data from the AGN shearing box simulations of \cite{jia16}. We also thank Shane Davis for his Monte Carlo code \citep{dav09}, which we modified to incorporate bulk Comptonization. This work was supported by NASA Astrophysics Theory Program grant NNX13AG61G and the International Space Science Institute (ISSI) in Bern. SH was supported by Japan JSPS KAKENH 15K05040 and the joint research
project of ILE, Osaka University. Numerical calculation of 110304a was
partly carried out on the Cray XT4 at CfCA, National Astronomical
Observatory of Japan, and  on SR16000 at YITP in Kyoto University.

%\onecolumn
\appendix
\section{Monte Carlo implementation of bulk Compton scattering}
\label{sec_MCdetails}
We incorporated bulk velocities into the Monte Carlo code used by \cite{dav09}, which is based on the statistically weighted photon packet method described in \cite{poz83}. Although the applications in this paper are nonrelativistic, we use exact Lorentz transforms. To test our code, we ran simulations with relativistic velocity fields and checked that spectra resulting from Lorentz transforming the emissivity were the same as spectra from simulations with a Lorentz-boosted field. We also ran simulations of Comptonization by divergenceless velocity fields and checked that the results were in agreement with the results of KB16.

The modifications we made in order to take bulk velocites into account are as follows. Photon packets are sampled from an emission function defined in the fluid frame, $\eta_0(\epsilon_0,{\bf \hat{\ell}_0})$, such as thermal brehmsstrahlung. The variables $\epsilon_0$ and ${\bf \hat{\ell}_0}$ denote the fluid frame photon energy and angle, respectively. Since the density grid is defined in the lab frame, we transform the density at a given point to the fluid frame before evaluating $\eta_0(\epsilon_0,{\bf \hat{\ell}_0})$. In this frame, the number of photons with energies between $\epsilon_0$ and $\epsilon_0 + d \epsilon_0$ within a solid angle $d {\bf \Omega_0}$ per unit time per unit volume is $f_0(\epsilon_0,{\bf \hat{\ell}_0}) d \epsilon_0 d {\bf \Omega_0} = [\eta_0(\epsilon_0,{\bf \hat{\ell}_0})/\epsilon_0] d \epsilon_0 d {\bf \Omega_0}$. The photon packet is then assigned a fluid frame statistical weight proportional to $f_0(\epsilon_0,{\bf \hat{\ell}_0})$. Lab frame energies and directions are calculated with standard Lorentz transforms, but calculating the correct lab frame statistical weight is more subtle. Since we want to sample from the lab frame photon number emissivity (i.e., per unit time, per unit volume) distribution $f(\epsilon,{\bf \hat{\ell}})$, 
\begin{align}
f(\epsilon,{\bf \hat{\ell}}) &= \frac{\eta(\epsilon,{\bf \hat{\ell}})}{\epsilon} \notag \\
&=  \left(\frac{\epsilon}{\epsilon_0}\right) \frac{\eta
_0(\epsilon_0,{\bf \hat{\ell}_0})}{\epsilon_0} \notag \\
&=  \left(\frac{\epsilon}{\epsilon_0}\right) f_0(\epsilon_0,{\bf \hat{\ell}_0}),
\end{align}
it may seem that the fluid frame statistical weight should be multiplied by $\epsilon / \epsilon_0$, but this is in fact incorrect. To see why, note that even without changing the statistical weight, simply boosting the energy and direction already results in a new distribution,
\begin{align}
f_0(\epsilon_0(\epsilon,{\bf \hat{\ell}}),{\bf \hat{\ell}_0}(\epsilon,{\bf \hat{\ell}}))\frac{\partial(\epsilon_0,{\bf \hat{\ell}_0})}{\partial(\epsilon,{\bf \hat{\ell}})},
\end{align}
which differs from the original distribution by the change of measure factor. Since the evaluation of this factor yields
\begin{align}
\frac{\partial(\epsilon_0,{\bf \hat{\ell}_0})}{\partial(\epsilon,{\bf \hat{\ell}})} = \frac{\epsilon}{\epsilon_0},
\end{align}
it so happens that the new distribution is already the lab frame photon number emissivity:
\begin{align}
f_0(\epsilon_0(\epsilon,{\bf \hat{\ell}}),{\bf \hat{\ell}_0}(\epsilon,{\bf \hat{\ell}}))\frac{\partial(\epsilon_0,{\bf \hat{\ell}_0})}{\partial(\epsilon,{\bf \hat{\ell}})} &= f_0(\epsilon_0(\epsilon,{\bf \hat{\ell}}),{\bf \hat{\ell}_0}(\epsilon,{\bf \hat{\ell}})) \frac{\epsilon}{\epsilon_0} \notag \\
&= f(\epsilon,{\bf \hat{\ell}}).
\end{align}
Therefore, the fluid frame and lab frame statistical weights are equal. Once a photon packet's lab frame energy, direction, and statistical weight are assigned, the method used to evolve it is in essence the same as in \cite{dav09}. Fluid frame parameters are self-consistently used in scattering events, and lab frame parameters are used to calculate changes in photon position between events. Fluid frame absorption coefficients are evaluated with densities transformed to the fluid frame.

\section{Derivation of the radiation pressure profile scaling}
\label{sec_pressure_scale}
The hydrostatic equilibrium equation is
\begin{align}
\frac{dP}{dz} = -\rho z \Omega^2_z,
\end{align}
where $\Omega_z$ is the vertical epicyclic frequency. The pressure profile is
\begin{align}
P(z) =& P_{\rm ph,in} + \Omega_z^2 \int_{z}^{z_{\rm ph}} \rho\left(z'\right)z' dz' \notag \\
=& P_{\rm ph,in} + \Omega_z^2 \left(\frac{\Sigma}{\Sigma_0} \right) \left(\frac{h}{h_0} \right)^{-1} \int_{z}^{z_{\rm ph}} \rho_0\left(h_0 z'/h\right)z' dz' \notag \\
=& P_{\rm ph,in} + \Omega_z^2 \left(\frac{\Sigma}{\Sigma_0} \right) \left(\frac{h}{h_0} \right)^{-1} \left( \int_{z}^{h z_{\rm ph,0}/h_0} \right. \notag \\
 &- \left. \int_{z_{\rm ph}}^{h z_{\rm ph,0}/h_0} \right) \rho_0\left(h_0 z'/h\right)z' dz',
\label{P2}
\end{align}
where the subscript ``ph'' denotes a value at the photosphere.
Therefore,
\begin{align}
P_0\left(h_0 z /h\right) =& P_{\rm ph,in,0} \notag \\
&+ \Omega_{z,0}^2 \left(\frac{h}{h_0}\right)^{-2} \int_{z}^{h z_{\rm ph,0}/h_0} \rho_0 \left( h_0 z'/h\right)z' dz'
\label{P3}
\end{align}
and
\begin{align}
P_0\left(h_0 z_{\rm ph} /h\right) =& P_{\rm ph,in,0} \notag \\
&+ \Omega_{z,0}^2 \left(\frac{h}{h_0}\right)^{-2} \int_{z_{\rm ph}}^{h z_{\rm ph,0}/h_0} \rho_0 \left( h_0 z'/h\right)z' dz'.
\label{P4}
\end{align}
Substitution of Eq. (\ref{P3}), Eq. (\ref{P4}), and Eq. (\ref{eq_hydro}) into Eq. (\ref{P2}) gives Eq. (\ref{eq_pressure_profile}). 

\section{Shearing box scalings in terms of flux, shear, and vertical epicyclic frequency}
\label{sec_shearing_scale}
The surface density scaling is
\begin{align}
\left(\frac{\Sigma}{\Sigma_0}\right) = \left(\frac{\alpha}{\alpha_0}\right)^{-1}
\left(\frac{\kappa}{\kappa_0}\right)^{-2}
\left(\frac{\Omega_z}{\Omega_{\rm z,0}}\right)^{2}
\left(\frac{\partial_x v_y}{\partial_x v_{y,0}}\right)^{-1}
\left(\frac{F}{F_0}\right)^{-1}.
\end{align}
The scale height scaling is
\begin{align}
\left(\frac{h}{h_0}\right) = \left(\frac{\kappa}{\kappa_0}\right) \left(\frac{\Omega_z}{\Omega_{z,0}}\right)^{-2}\left(\frac{F}{F_0}\right).
\end{align}
The density profile scaling is
\begin{align}
\rho\left(z\right) =& \left(\frac{\alpha}{\alpha_0}\right)^{-1}
\left(\frac{\kappa}{\kappa_0}\right)^{-3}
\left(\frac{\Omega_z}{\Omega_{\rm z,0}}\right)^{4}
 \left(\frac{\partial_x v_y}{\partial_x v_{y,0}}\right)^{-1} \notag \\
&\left(\frac{F}{F_0}\right)^{-2}
\rho_0\left(h_0 z/h\right).
\end{align}
The scalings for the pressure and gas temperature profiles are given by Eqs. (\ref{eq_pressure_profile}), (\ref{eq_Tgas_profile_1}), and (\ref{eq_Tgas_profile_2}). The turbulent velocity profile scaling is
\begin{align}
v(z) =& \left(\frac{\alpha}{\alpha_0}\right)^{1/2}
 \left(\frac{\beta}{\beta_0}  \right)^{1/2}
\left(\frac{\kappa}{\kappa_0}\right) \left(\frac{\Omega_z}{\Omega_{z,0}}\right)^{-1} \notag \\
&\left(\frac{F}{F_0} \right) v_0(h_0 z/h).
\label{eq_v_turb_AC}
\end{align}
The shear velocity profile scaling is
\begin{align}
v_{\rm s}\left(x\right) = \left(\frac{\kappa}{\kappa_0}\right)
 \left(\frac{\Omega_z}{\Omega_{\rm z,0}} \right)^{-2} \left(\frac{\partial_x v_y}{\partial_x v_{y,0}} \right) \left(\frac{F}{F_0} \right) v_{\rm s,0}\left(h_0 x/h\right).
\end{align}

\section{Scalings for flux, shear and vertical epicyclic frequency in terms of radius, mass, and accretion rate}
\label{sec_parameter_scale}

\subsection{Newtonian scalings}
\label{sec_Newt_1}
Let $M$ and $\dot{M}$ be the mass and mass accretion rate, respectively. Define $r = R/R_{\rm g}$ and $\dot{m} = \dot{M}/\dot{M}_{\rm Edd}$. We also define
\begin{align}
\eta = \frac{1}{2r_{\rm in}}.
\end{align}
The flux, derived from energy and angular momentum conservation, is given by \cite{ago00} Eq. (11):
\begin{align}
F = \frac{3GM\dot{M}}{8\pi R^3} \left(1-\sqrt{r_{\rm in}/r} + \left(\sqrt{r_{\rm in}/r}\right) r_{\rm in} \Delta \epsilon \right),
\end{align}
where $\Delta \epsilon$ is the change in efficiency due to a non-zero stress-free inner boundary condition. The flux scaling is
\begin{align}
\left(\frac{F}{F_0}\right) =& \left(\frac{r}{r_0} \right)^{-3} \left(\frac{M}{M_0} \right)^{-1} \left(\frac{\dot{m}}{\dot{m}_0} \right) \left(\frac{\eta + \Delta \epsilon}{\eta_0 + \Delta \epsilon_0} \right)^{-1} \notag \\
&\left(\frac{ 1-\sqrt{r_{\rm in}/r} + \left(\sqrt{r_{\rm in}/r}\right) r_{\rm in} \Delta \epsilon }{ 1-\sqrt{r_{\rm in,0}/r_0} + \left(\sqrt{r_{\rm in,0}/r_0}\right) r_{\rm in,0} \Delta \epsilon_0 }\right).
\end{align}
The vertical epicyclic frequency is
\begin{align}
\Omega_z = \sqrt{\frac{GM}{R^3}}.
\end{align}
The scaling for the vertical epicyclic frequency is
\begin{align}
\left(\frac{\Omega_z}{\Omega_{z,0}}\right) = \left(\frac{M}{M_0}\right)^{-1} \left(\frac{r}{r_0} \right)^{-3/2}.
\end{align}
The strain rate is
\begin{align}
\partial_x v_y = \frac{3}{2} \sqrt{\frac{GM}{R^3}}.
\end{align}
The strain rate scaling is
\begin{align}
\left(\frac{\partial_x v_y}{\partial_x v_{y,0}}\right)
 = \left(\frac{M}{M_0}\right)^{-1} \left(\frac{r}{r_0} \right)^{-3/2} .
\end{align}

\subsection{Kerr scalings}
\label{sec_kerr_1}
Let $M$ and $\dot{M}$ be the mass and mass accretion rate, respectively. Let $R$ be the Boyer-Linquist radial coordinate and $a$ be the dimensionalized spin parameter. Define $r = R/R_{\rm g}$ and $\dot{m} = \dot{M}/\dot{M}_{\rm Edd}$. The expressions for $A$, $B$, $C$, $D$, and $E$ are given by \cite{rif95} (hereafter, RH95) Eq. (6). In terms of the dimensionalized variables, they are
\begin{align}
A = 1 - \frac{2}{r} + \frac{a^2}{r^2},
\end{align}
\begin{align}
B = 1 - \frac{3}{r} + \frac{2 a}{r^{3/2}},
\end{align}
\begin{align}
C = 1 - \frac{4a}{r^{3/2}} + \frac{3 a^2}{r^2},
\end{align}
\begin{align}
D = \frac{1}{2 \sqrt{r}} \int_{r_{\rm in}}^{r} \frac{x^2 - 6x + 8a \sqrt{x} - 3a^2}{\sqrt{x} \left(x^2 - 3x + 2a \sqrt{x} \right)}dx,
\end{align}
\begin{align}
E = 1 - \frac{6}{r} + \frac{8a}{r^{3/2}} - \frac{3a^2}{r^2},
\end{align}
where $r_{\rm in}$ is given by $E(r_{\rm in}) = 0$. We also define 
\begin{align}
\eta = 1 - \left(1-\frac{2}{3 r_{\rm in}}\right)^{1/2},
\end{align}
the efficiency parameter assuming a stress-free inner boundary condition.
The flux is given by the thermal equilibrium equation, RH95 Eq. (19), modified by the non-zero stress inner boundary term in Agol \& Krolik (2000) Eq. (8):
\begin{align}
F &= \frac{3\dot{M}M}{8 \pi R^3} B^{-1} \left(r_{\rm in}^{3/2} B(r_{\rm in})^{1/2} \Delta \epsilon r^{-1/2} + D \right),
\end{align}
where $\Delta \epsilon$ is the change in efficiency due to a non-zero stress-free inner boundary condition. The flux scaling is
\begin{align}
\left(\frac{F}{F_0}\right) =& \left(\frac{r}{r_0} \right)^{-3} \left(\frac{M}{M_0} \right)^{-1} \left(\frac{\dot{m}}{\dot{m}_0} \right) \left(\frac{\eta + \Delta \epsilon}{\eta_0 + \Delta \epsilon_0} \right)^{-1} \notag \\
&\left(\frac{B}{B_0}\right)^{-1} 
\left(\frac{r_{\rm in}^{3/2} B(r_{\rm in})^{1/2} \Delta \epsilon r^{-1/2} + D }{r_{\rm in,0}^{3/2} B(r_{\rm in,0})^{1/2} \Delta \epsilon_0 r_0^{-1/2} + D_0 }\right).
\end{align}
The vertical epicyclic frequency, inferred from RH95 Eq. (12), is
\begin{align}
\Omega_z = \sqrt{\frac{GM}{R^3}CB^{-1}}.
\end{align}
The scaling for the vertical epicyclic frequency is
\begin{align}
\left(\frac{\Omega_z}{\Omega_{z,0}}\right) = \left(\frac{M}{M_0}\right)^{-1} \left(\frac{r}{r_0} \right)^{-3/2} \left(\frac{C}{C_0}\right)^{1/2}\left(\frac{B}{B_0}\right)^{-1/2}.
\end{align}
The strain rate, inferred from RH95 Eq. (14), is
\begin{align}
\partial_x v_y = \frac{3}{2} \sqrt{\frac{GM}{R^3}} A B^{-1}.
\end{align}
The strain rate scaling is
\begin{align}
\left(\frac{\partial_x v_y}{\partial_x v_{y,0}}\right) = \left(\frac{M}{M_0}\right)^{-1} \left(\frac{r}{r_0} \right)^{-3/2} \left(\frac{A}{A_0}\right) \left(\frac{B}{B_0}\right)^{-1}.
\end{align}

\section{Shearing box scalings in terms of radius, mass, and accretion rate}
\label{sec_final_scale}
\subsection{Newtonian scalings}
In this section we substitute the results of Appendix \ref{sec_Newt_1} into the results of Appendix \ref{sec_shearing_scale}. The density profile scaling is
\begin{align}
\rho\left(z\right) =& \left(\frac{\alpha}{\alpha_0}\right)^{-1}
\left(\frac{\kappa}{\kappa_0}\right)^{-3}
\left(\frac{r}{r_0} \right)^{3/2} \left(\frac{M}{M_0} \right)^{-1}  \notag\\
 &\left(\frac{\dot{m}}{\dot{m}_0} \right)^{-2}
 \left(\frac{\eta + \Delta \epsilon}{\eta_0 + \Delta \epsilon_0} \right)^{2} \notag \\
 &\left(\frac{ 1-\sqrt{r_{\rm in}/r} + \left(\sqrt{r_{\rm in}/r}\right) r_{\rm in} \Delta \epsilon }{ 1-\sqrt{r_{\rm in,0}/r_0} + \left(\sqrt{r_{\rm in,0}/r_0}\right) r_{\rm in,0} \Delta \epsilon_0 }\right)^{-2} \rho_0 \left(h_0 z/h\right).
\end{align}
The pressure profile scaling is given by Eq. (\ref{eq_pressure_profile}), and the gas temperature profile scaling is given by Eqs. (\ref{eq_Tgas_profile_1}), (\ref{eq_Tgas_ph}), (\ref{eq_Tgas_profile_2}), where
\begin{align}
\left(\frac{P_{\rm c}}{P_{\rm c,0}}\right) = \left(\frac{\alpha}{\alpha_0}\right)^{-1}
\left(\frac{\kappa}{\kappa_0}\right)^{-1}
 \left(\frac{r}{r_0} \right)^{-3/2} \left(\frac{M}{M_0} \right)^{-1},
\end{align}
and
\begin{align}
P_{\text{ph}} = &\left(\frac{f_{\text{cor}}}{f_{\text{cor},0}}\right)^{4} \left(\frac{r}{r_0}\right)^{-3} \left(\frac{M}{M_0}\right)^{-1} \left(\frac{\dot{m}}{\dot{m}_0}\right)\left(\frac{\eta + \Delta \epsilon}{\eta_0 + \Delta \epsilon_0} \right)^{-1} \notag \\
&\left(\frac{ 1-\sqrt{r_{\rm in}/r} + \left(\sqrt{r_{\rm in}/r}\right) r_{\rm in} \Delta \epsilon }{ 1-\sqrt{r_{\rm in,0}/r_0} + \left(\sqrt{r_{\rm in,0}/r_0}\right) r_{\rm in,0} \Delta \epsilon_0 }\right) P_{\text{ph},0}.
\end{align}
The turbulent velocity profile scaling is
\begin{align}
v(z) = &\left( \frac{\alpha}{\alpha_0} \right)^{1/2}
\left( \frac{\beta}{\beta_0} \right)^{1/2}
\left(\frac{\kappa}{\kappa_0}\right)
 \left( \frac{r}{r_0} \right)^{-3/2}
 \left( \frac{\dot{m}}{\dot{m}_0} \right) \notag \\
 &\left(\frac{\eta + \Delta \epsilon}{\eta_0 + \Delta \epsilon_0} \right)^{-1} \notag \\
 & \left(\frac{ 1-\sqrt{r_{\rm in}/r} + \left(\sqrt{r_{\rm in}/r}\right) r_{\rm in} \Delta \epsilon }{ 1-\sqrt{r_{\rm in,0}/r_0} + \left(\sqrt{r_{\rm in,0}/r_0}\right) r_{\rm in,0} \Delta \epsilon_0 }\right)
 v_0(h_0 z/h).
\end{align}
The shear velocity profile scaling is
\begin{align}
v_{\rm s}\left(x\right)  = &\left(\frac{\kappa}{\kappa_0}\right)
\left(\frac{r}{r_0}\right)^{-3/2} \left(\frac{\dot{m}}{\dot{m}_0}\right) \left(\frac{\eta + \Delta \epsilon}{\eta_0 + \Delta \epsilon_0} \right)^{-1} \notag \\
&\left(\frac{ 1-\sqrt{r_{\rm in}/r} + \left(\sqrt{r_{\rm in}/r}\right) r_{\rm in} \Delta \epsilon }{ 1-\sqrt{r_{\rm in,0}/r_0} + \left(\sqrt{r_{\rm in,0}/r_0}\right) r_{\rm in,0} \Delta \epsilon_0 }\right) v_{\rm s,0}\left(h_0 x/h\right).
\end{align}
The surface density profile scaling is
\begin{align}
\left(\frac{\Sigma}{\Sigma_0}\right) =& \left(\frac{\alpha}{\alpha_0}\right)^{-1}
\left(\frac{\kappa}{\kappa_0}\right)^{-2}
\left(\frac{r}{r_0}\right)^{3/2}
\left(\frac{\dot{m}}{\dot{m}_0}\right)^{-1} \notag \\
&\left(\frac{ 1-\sqrt{r_{\rm in}/r} + \left(\sqrt{r_{\rm in}/r}\right) r_{\rm in} \Delta \epsilon }{ 1-\sqrt{r_{\rm in,0}/r_0} + \left(\sqrt{r_{\rm in,0}/r_0}\right) r_{\rm in,0} \Delta \epsilon_0 }\right)^{-1},
\end{align}
and the scale height scaling is
\begin{align}
\left(\frac{h}{h_0}\right) =& \left(\frac{\kappa}{\kappa_0}\right)
\left(\frac{M}{M_0}\right)
\left(\frac{\dot{m}}{\dot{m_0}}\right)
\left(\frac{\eta + \Delta \epsilon}{\eta_0 + \Delta \epsilon_0} \right)^{-1}\notag \\
 &\left(\frac{ 1-\sqrt{r_{\rm in}/r} + \left(\sqrt{r_{\rm in}/r}\right) r_{\rm in} \Delta \epsilon }{ 1-\sqrt{r_{\rm in,0}/r_0} + \left(\sqrt{r_{\rm in,0}/r_0}\right) r_{\rm in,0} \Delta \epsilon_0 }\right).
\end{align}

\subsection{Kerr scalings}
In this section we substitute the results of Appendix \ref{sec_kerr_1} into the results of Appendix \ref{sec_shearing_scale}. The density profile scaling is
\begin{align}
\rho\left(z\right) = &\left(\frac{\alpha}{\alpha_0}\right)^{-1}
\left(\frac{\kappa}{\kappa_0}\right)^{-3}
\left(\frac{r}{r_0} \right)^{3/2} \left(\frac{M}{M_0} \right)^{-1} \left(\frac{\dot{m}}{\dot{m}_0} \right)^{-2} \notag \\
&\left(\frac{\eta + \Delta \epsilon}{\eta_0 + \Delta \epsilon_0} \right)^{2} \left(\frac{A}{A_0}\right)^{-1} \left(\frac{B}{B_0}\right) \left(\frac{C}{C_0}\right)^{2} \notag\\
&\left(\frac{r_{\rm in}^{3/2} B(r_{\rm in})^{1/2} \Delta \epsilon r^{-1/2} + D}{r_{{\rm in},0}^{3/2} B(r_{{\rm in},0})^{1/2} \Delta \epsilon_0 r_0^{-1/2} + D_0}\right)^{-2} \rho_0 \left(h_0 z/h\right).
\end{align}
The pressure profile scaling is given by Eq. (\ref{eq_pressure_profile}), and the gas temperature profile scaling is given by Eqs. (\ref{eq_Tgas_profile_1}), (\ref{eq_Tgas_ph}), (\ref{eq_Tgas_profile_2}), where
\begin{align}
\left(\frac{P_{\rm c}}{P_{\rm c,0}}\right) =& \left(\frac{\alpha}{\alpha_0}\right)^{-1}
\left(\frac{\kappa}{\kappa_0}\right)^{-1}
 \left(\frac{r}{r_0} \right)^{-3/2}
 \left(\frac{M}{M_0} \right)^{-1} \notag \\
 &\left(\frac{A}{A_0}\right)^{-1} \left(\frac{C}{C_0}\right),
\end{align}
and
\begin{align}
P_{\text{ph}} = &\left(\frac{f_{\text{cor}}}{f_{\text{cor},0}}\right)^{4} \left(\frac{r}{r_0}\right)^{-3} \left(\frac{M}{M_0}\right)^{-1} \left(\frac{\dot{m}}{\dot{m}_0}\right)\left(\frac{\eta + \Delta \epsilon}{\eta_0 + \Delta \epsilon_0} \right)^{-1} \notag \\
&\left(\frac{B}{B_0} \right)^{-1}\left(\frac{r_{\rm in}^{3/2} B(r_{\rm in})^{1/2} \Delta \epsilon r^{-1/2} + D}{r_{{\rm in},0}^{3/2} B(r_{{\rm in},0})^{1/2} \Delta \epsilon_0 r_0^{-1/2} + D_0 } \right) P_{\text{ph},0}.
\end{align}
The turbulent velocity profile scaling is
\begin{align}
v(z) = &\left( \frac{\alpha}{\alpha_0} \right)^{1/2}
\left( \frac{\beta}{\beta_0} \right)^{1/2}
\left(\frac{\kappa}{\kappa_0}\right)
 \left( \frac{r}{r_0} \right)^{-3/2} 
 \left( \frac{\dot{m}}{\dot{m}_0} \right) \notag \\
 &\left(\frac{\eta + \Delta \epsilon}{\eta_0 + \Delta \epsilon_0} \right)^{-1}
 \left(\frac{B}{B_0}\right)^{-1/2} 
  \left(\frac{C}{C_0}\right)^{-1/2} \notag \\ 
 & 
 \left(\frac{r_{\rm in}^{3/2} B(r_{\rm in})^{1/2} \Delta \epsilon r^{-1/2} + D}{r_{{\rm in},0}^{3/2} B(r_{{\rm in},0})^{1/2} \Delta \epsilon_0 r_0^{-1/2} + D_0}\right) v_0(h_0 z/h).
\end{align}
The shear velocity profile scaling is
\begin{align}
v_{\rm s}\left(x\right)  = &\left(\frac{\kappa}{\kappa_0}\right)
\left(\frac{r}{r_0}\right)^{-3/2} \left(\frac{\dot{m}}{\dot{m}_0}\right) \left(\frac{\eta + \Delta \epsilon}{\eta_0 + \Delta \epsilon_0} \right)^{-1} \notag \\
 &\left(\frac{A}{A_0}\right) \left(\frac{B}{B_0}\right)^{-1} \left(\frac{C}{C_0}\right)^{-1} \notag \\
&\left(\frac{r_{\rm in}^{3/2} B(r_{\rm in})^{1/2} \Delta \epsilon r^{-1/2} + D}{r_{{\rm in},0}^{3/2} B(r_{{\rm in},0})^{1/2} \Delta \epsilon_0 r_0^{-1/2} + D_0}\right) v_{\rm s,0}\left(h_0 x/h\right).
\end{align}
The surface density scaling is
\begin{align}
\left(\frac{\Sigma}{\Sigma_0}\right) =& \left(\frac{\alpha}{\alpha_0}\right)^{-1}
\left(\frac{\kappa}{\kappa_0}\right)^{-2}
\left(\frac{r}{r_0}\right)^{3/2}
\left(\frac{\dot{m}}{\dot{m}_0}\right)^{-1}
\left(\frac{A}{A_0}\right)^{-1} \notag \\
&\left(\frac{B}{B_0}\right)
\left(\frac{C}{C_0}\right)
\left(\frac{r_{\rm in}^{3/2} B(r_{\rm in})^{1/2} \Delta \epsilon r^{-1/2} + D}{r_{{\rm in},0}^{3/2} B(r_{{\rm in},0})^{1/2} \Delta \epsilon_0 r_0^{-1/2} + D_0}\right)^{-1},
\end{align}
and the scale height scaling is
\begin{align}
\left(\frac{h}{h_0}\right) =& \left(\frac{\kappa}{\kappa_0}\right)
\left(\frac{M}{M_0}\right)\left(\frac{\dot{m}}{\dot{m_0}}\right) \left(\frac{\eta + \Delta \epsilon}{\eta_0 + \Delta \epsilon_0} \right)^{-1}
\left(\frac{C}{C_0}\right)^{-1} \notag \\
&\left(\frac{r_{\rm in}^{3/2} B(r_{\rm in})^{1/2} \Delta \epsilon r^{-1/2} + D}{r_{{\rm in},0}^{3/2} B(r_{{\rm in},0})^{1/2} \Delta \epsilon_0 r_0^{-1/2} + D_0}\right).
\end{align}

\end{document}